\documentclass[a4paper,10pt]{article}

\usepackage[T1]{fontenc} 
\usepackage{fixltx2e} 
\usepackage{multirow}
\usepackage{epsfig}
\usepackage{rotating}
\usepackage{amsmath}
\usepackage{amssymb}
\usepackage{psboxit}
\usepackage[tight]{subfigure}
\usepackage{rotating}
\usepackage{euscript}
\usepackage{graphics}
\usepackage{txfonts}
\usepackage[usenames]{color}
\usepackage{figlatex}
\usepackage[colorlinks=false]{hyperref}
\usepackage[normalem]{ulem}
\usepackage{color}
\def\foofaaXSpace{%
 \ifx\foofaaletfoofaatoken\foofaasptoken\else%
  \ifx\foofaaletfoofaatoken\egroup\else%
   \ifx\foofaaletfoofaatoken\/\else%
    \ifx\foofaaletfoofaatoken\ \else%
     \ifx\foofaaletfoofaatoken\,\else%
      \ifx\foofaaletfoofaatoken~\else%
       \ifx\foofaaletfoofaatoken.\else%
        \ifx\foofaaletfoofaatoken!\else%
         \ifx\foofaaletfoofaatoken,\else%
          \ifx\foofaaletfoofaatoken:\else%
           \ifx\foofaaletfoofaatoken;\else%
            \ifx\foofaaletfoofaatoken?\else%
             \ifx\foofaaletfoofaatoken'\else%
              \ifx\foofaaletfoofaatoken`\else%
               \ifx\foofaaletfoofaatoken)\else%
                \ifx\foofaaletfoofaatoken]\else%
                 \ifx\foofaaletfoofaatoken\}\else%
                  \ifx\foofaaletfoofaatoken-\else%
                   \ifx\foofaaletfoofaatoken\foofaa\else%
                        \ifmmode\relax\else\space\fi%
                   \fi%
                  \fi%
                 \fi%
                \fi%
               \fi%
              \fi%
             \fi%
            \fi%
           \fi%
          \fi%
         \fi%
        \fi%
       \fi%
      \fi%
     \fi%
    \fi%
   \fi%
  \fi%
 \fi%
}

\DeclareRobustCommand\XSpace{\futurelet\foofaaletfoofaatoken\foofaaXSpace}

\def\hi{{\it h-index}\XSpace}

\def\chi{{\it contemporary h-index}\XSpace}

\begin{document}

\title{Identification of \\
Influential Scientists vs. Mass Producers \\
by the Perfectionism Index}
\author{Antonis Sidiropoulos$^{1,3}$ \ \ Dimitrios Katsaros$^2$
\thanks{Corresponding author: Dimitrios Katsaros ({\tt dkatsar@inf.uth.gr})} 
\ \ Yannis Manolopoulos$^1$ \\ \\
$^1$Department of Informatics, \\ Aristotle University, Thessaloniki, Greece \\ \\
$^2$Department of Computer \& Communications Engineering, \\ University of Thessaly, Volos, Greece \\ \\ 
$^3$Department of Information Technology, \\ 
Alexander Technological Educational Institute of Thessaloniki, \\ Thessaloniki, Greece \\ \\
\tt \{asidirop,manolopo\}@csd.auth.gr, dkatsar@inf.uth.gr
}

\maketitle

\begin{abstract}
The concept of \hi has been proposed to easily assess a researcher's performance with 
a single number. However, by using only this number, we lose significant information 
about the distribution of citations per article in an author's publication list. 
In this article, we study an author's citation curve and we define two new areas 
related to this curve. We call these ``penalty areas", since the greater 
they are, the more an author's performance is penalized. We exploit these areas 
to establish new indices, namely PI and XPI, aiming at categorizing researchers in 
two distinct categories: ``influentials" and ``mass producers"; the former category 
produces articles which are (almost all) with high impact, and the latter category 
produces a lot of articles with moderate or no impact at all. Using data from 
Microsoft Academic Service, we evaluate the merits mainly of PI as a useful tool 
for scientometric studies. We establish its effectiveness into separating the 
scientists into influentials and mass producers; we demonstrate its robustness 
against self-citations, and its uncorrelation to traditional indices. Finally, 
we apply PI to rank prominent scientists in the areas of databases, networks and 
multimedia, exhibiting the strength of the index in fulfilling its design goal.

\end{abstract}

\section{Introduction}
\label{sec:intro}

The \hi has been a well honored concept since it was proposed by Jorge 
Hirsch~\cite{Hirsch-PNAS05}. A lot of variations have been proposed in the 
literature, see for instance the references within~\cite{Alonso-JNL-Informetrics09}. 
Many efforts enhanced the original \hi by taking into account 
age-related issues~\cite{Sidiropoulos-Scientometrics07}, 
multi-authorship~\cite{Hirsch-Scientometrics10}, fractional 
citation counting~\cite{Katsaros-JASIST09}, the highly cited 
articles~\cite{Egghe-Scientometrics06}. Other works explored its 
predictive capabilities~\cite{Hirsch-PNAS07}, its robustness to 
self-citations~\cite{Schreiber-EPL07}, etc. Some of the proposals 
have been implemented in commercial and free software, such as 
Matlab\footnote{\url{http://www.mathworks.de/matlabcentral/fileexchange/28161-bibliometrics-the-art}\\\url{-of-citations-indices}} 
and the Publish or Perish software~\footnote{\url{http://www.harzing.com/pop.htm}}.

\begin{figure}[!ht]
\centerline{\psfig{figure=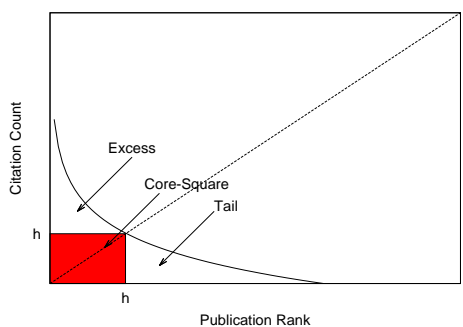,clip=,width=2.5in}}
\caption{Citation curve depicting the excess, core and tail areas.}
\label{example_plain_fig}
\end{figure}

Even though there are several hundreds of articles developing variations to the 
original \hi, there is notably little research on making a better and deeper 
exploitation of the ``primitive" information that is carried by the citation 
curve itself and by its intersection with the $45^o$ line defining the \hi. 
The projection of the intersection point on the axes creates three areas that 
were termed in~\cite{Rousseau-Science-Focus06}, \cite{Ye-Scientometrics10}, 
and~\cite{Zhang-PLOSOne09} as the {\it h-core-square} area\footnote{In the sequel of the article 
for the sake of simplicity, we use the term h-core and h-core-square interchangeably.   
}, the {\it tail} area and the {\it excess} area (see 
Figure~\ref{example_plain_fig}). The core area is a square of size $h$ (depicted 
by grey color in the figure), includes $h^{2}$ citations and it is also called Durfee 
square area~\cite{Anderson-Scientometrics08}; the area that lies to the right of 
the core area is the tail or {\it lower area}, whereas the area above the 
core area is the excess or {\it upper} or $e^2$ area~\cite{Zhang-PLOSOne09}. 
Both the absolute and the relative sizes of these areas carry significant information. 
The absolute size of the excess and core areas were directly used for the definition 
of e-index and \hi; part of the absolute size of the tail area was 
used in~\cite{Garcia-JNL-Informetrics12} to create a vector of h-indices; the 
relative size of the core to the tail area (without taking into account the 
tail length) was used in~\cite{Ye-Scientometrics10} for similar purposes, etc. (
For a complete review of the relevant bibliography cf.\ section~\ref{sec-relevant}.) 
The common characteristic of all these works is that they develop indices to ``break 
ties", i.e., to differentiate between scientists with equal h-indices, and thus 
address the isohindex problem.

We believe that the latent information carried by these areas is not adequately 
explored, and most significantly, it can be used in a different way, not as a plain 
tie-breaker, but as a ``first-class" citizen in the scientometric indices ecosystem. 

Rosenberg~\cite{Rosenberg-TR11} took the first step towards this goal; 
he provided a {\it qualitative characterization} for the scientists with many 
citations in the upper area and a few citations in the tail area by referring to them as 
{\it perfectionists}. He referred to the authors with few citations in the upper area 
and many citations in the tail area as {\it mass producers}, since they have a 
lot of publications but mostly of low impact. Finally, he referred to the rest of the 
scientists as the {\it prolific} ones. The origin of this terminology is quite old; 
it was proposed in~\cite{Cole-ASR67}, and subsequently studied further in~\cite{Feist-CRJ97}.

Motivated by Rosenberg's classification scheme, we pose the following question: 
``Can we develop a quantitative methodology for identifying those scientists who 
are truly ``laconic" and constantly influential compared to those who produce 
a mass of papers but relatively few of them contribute to their \hi?"

In the present paper, we will present a methodology and an easy calculated criterion 
to categorize a scientist in one of two distinct categories: either an author is 
a ``mass producer", i.e., he has authored many papers with relatively few citations 
or he is ``influential", i.e., most of his papers have an impact because they 
have received a significant number of citations. This methodology will indirectly 
highlight the ``attitude" towards publishing. Some scientists are acting in a laconic 
way, in the sense that they are not fond of having published ``half-baked" articles 
that are soon superseded by mature and extended versions of their work.  
Others develop their work in a slow and incremental way, publishing their ideas in 
a step-by-step fashion producing a lot of moderate impact articles until they hit 
the big contribution. This attitude may be due to other reasons as well, e.g., the 
pressure to present published articles as deliverables to a project. In any case, 
it is not the purpose of the present article to discover and explain those reasons. 
The sole purpose of our work is to develop metrics that can be used complementary to 
the traditional ones such as the \hi, in order to separate the steadily influential 
authors from the mass producers. At this point we need to emphasize that the concept of 
``influential" scientists we develop here is not related to the notion of influential 
nodes in a social network of actors as considered in~\cite{Basaras-IEEE-Computer13}.

The area of scientometric performance indicators is very rich, and it
is continuously flourishing. Vinkler in~\cite{Vincler-JASIST11}
provides a brief classification of the traditional and modern
scientometric
indicators explaining their virtues and shortcomings; it is shown
there that Hirsch index is not the only indicator that combines impact
and quantity, but $\pi$-index~\cite{Vincler-JoIS09} which introduced
the concept of ``elite set" is another competitor of it.
Nevertheless, in this article we use Hirsch index as a basis to
expose our ideas claiming that neither Hirsch index is the 'best'
one nor that our core methodology applies exclusively to it.
We are strongly
confident that our ideas can be applied also to the family of indices
based on the Impact Factor by penalizing those journals that publish
articles which accumulate far less citations than the Impact Factor
of the journal they appear in.

The rest of the article is organized as follows. In the next section we will 
present the relevant literature and then define two new areas in the citation 
curve. Based on these two new areas, we will establish two new metrics for evaluating 
the performance of authors in terms of impact. In Section~\ref{sec:experiments} 
we will present our datasets, which were built by extracting data from the 
Microsoft Academic Search database, and analyze these data to view the dataset 
characteristics. Subsequently, we will present the distributions of our new metrics for 
the above datasets and compare them with other metrics proposed in 
the literature. Finally, in Section~\ref{sec:results} we will present some of the 
resulting ranking tables based on the new metrics and \hi. Section~\ref{sec:conclusion} 
will conclude the article.

\subsection{Motivation and contributions}
\label{subsec-motive}

During the latest  years an abundance of scientometric indices have been published to 
evaluate the academic merit of a scientist. Despite the debate around the usefulness 
of any index in general, they remain an indispensable part of the evaluation 
process of a scientist's academic merit. The ideas behind the \hi philosophy was so 
influential, that the vast majority of the proposed indices are about some variant or 
extension of the \hi itself. Despite the wealth and sophistication of the proposed 
indices, we argue that the relevant literature did not strive for an {\it holistic} 
consideration of the information carried by the citation curve and by the $45^o$ 
line. In the next paragraph we will present the motivating idea with a simple example.

Let us consider author $A$ who has published~13 articles, and author 
$B$ who has published~24 articles with citation distributions 
$\{29,24,20,17,15,14,13,12,11,10,9,3,0\}$ and 
$\{29,24,20,17,15,14,13,12,11,10,2,1,1,1,1,1,1,1,1,1,1,0,0,0\}$ respectively. 
Both authors have the same ``macroscopic" characteristics in terms of the number 
of citations, i.e., they both have the same total number of citations, identical 
core areas and h-indices equal to~10, identical excess areas with~65 citations 
there, and the same number of citations in the tail area, namely 12. However, author 
$A$ has only 3 articles in his tail area, whereas author~$B$ has 14 articles.

At a first glance, we can simply use the number of articles in the tail as a 
tie-breaker to differentiate between the two authors, and characterize the first one 
as constantly ``influential", and the second one as a ``mass producer". But, how can 
we capture the fact that, in the short term, the first author's  \hi is more likely to 
increase.  At the same time, 
we need a way to describe -- actually, to penalize -- the second author for this long 
and lightweight tail. Starting from these questions we will define the {\it penalty 
areas} and then develop the respective indices. In this article, we do not consider 
temporal issues, e.g., the time of publication of the articles in the tail area; 
such issues are part of our on-going work. Specifically, the article makes the 
following contributions:
\begin{itemize}
\item 
It defines two areas to quantify the fact that some authors publish articles  which 
eventually do not have analogous impact with those that contribute to their \hi.
\item 
It develops two new perfectionism indices taking into account the size of the penalty areas. There are the $PI$ and $XPI$ indices, which are 
statistically uncorrelated to the \hi, 
thus proving that they measure something that is different from what the \hi measures. 
\item 
Using these indices, it proposes a filter to separate the authors into influential 
ones and mass producers. This filter partitions the authors irrespectively of their 
\hi, i.e., it can classify two authors as influentials, even if the values of their 
h-indices are significantly different. This two-way segmentation of the scientists 
is a significant departure from the earlier, rich classification 
schemes~\cite{Cole-ASR67,Feist-CRJ97}, since with a {\it single} integer number 
and its sign (plus or minus) it can provide rankings, contrary to intuitive mapping 
schemes such as that in~\cite{Zhang-NatureSR13}. 
\item 
It provides a thorough investigation of the indices against the \hi for three 
datasets retrieved by the Microsoft's Academic Search.
\end{itemize} 

At this point, we need to emphasize again that the proposed indices are neither 
a substitute for any of the already existing metrics nor a tool for 
identifying ``bad" publishing behaviour. They are one more tool in the indices 
toolbox of someone who wishes to capture the multi-dimensional facet of a 
scientist's performance.

\section{Relevant work}
\label{sec-relevant}

The original article by Hirsch~\cite{Hirsch-PNAS05} created a huge wave of 
proposals for indices attempting to capture the academic performance of a scientist. 
It is characteristic that at the time of writing the present article, the \hi's 
article had more than 3850 citations in Google Scholar. Since the focus of the 
present manuscript is not about the \hi in general, but about the exploitation of 
the information in the {\it tail area}, we will survey only the articles relevant 
to the usage and mining of that part of the citation. 

Ye and Rousseau~\cite{Ye-Scientometrics10} studied the evolution of tail-core 
ratio as a function of time, and later extended their study 
in~\cite{Liu-JNL-Informetrics13} including a few more ratios among the three areas. 
Similar in spirit is the work reported in~\cite{Chen-JNL-Informetrics13}, 
which examines variations of the ratios across scientific disciplines. 
Baum~\cite{Baum-TR-12} introduced the ratio (the relative citedness) of the few, 
highly-cited articles in a journal's h-core and the many, infrequently-cited 
articles in its h-tail as a way to improve journals' Impact Factors.

Having as motive to consider each and every citation under the {\it whole} 
citation curve (and therefore under the tail area as well) , Anderson et al.\ 
\cite{Anderson-Scientometrics08} proposed a fractional citation counting scheme 
based on Ferrers graphs. Later, Franceschini and 
Maisano~\cite{Franceschini-JNL-Informetrics10} recognized the weaknesses of that 
scheme and proposed the Citation Triad method; both indices are striving to exploit 
the information under the whole citation curve in a way that creates a strictly 
monotonic (increasing) index for every new citation added to the curve, which is 
completely different to what we propose. 

A kind of ``quantization" scheme for the citation curve and the creation of multiple 
Durfee squares under that curve was proposed in~\cite{Garcia-JNL-Informetrics12}. 
The output of that method was a vector (i.e., multiple h-indexes) as a measure of the 
scientific performance. However, the method simply transformed the task of comparing 
different citation curves into the problem of comparing vectors, without setting 
clear rules. A study of the contribution of the excess, core and tail areas to the 
entire citation curve was performed in~\cite{Bornmann-JNL-Informetrics10} proving 
that this
contribution varies across scientists. The study provided also a regression model 
for determining the most visible article of a scientist. The position of the 
centroids of the core and tail area was used in~\cite{Kuan-JNL-Informetrics11} 
as an index for comparing different scientists providing only straightforward 
characterizations for high-low impact and productivity. Along these lines of 
research, Zhang~\cite{Zhang-NatureSR13} proposed a triangle mapping technique to map 
the three percentages (of the excess, core and tail area) of these citations onto a 
point within a regular triangle; by viewing the distribution of the mapping points, 
different shapes of citation curves can be studied in a perceivable form. The work 
described in~\cite{Dorta-Scientometrics11} sought selective and large producers 
considering only a part of the excess and a part of the tail area, thus again 
neglecting a part of the tail area which carries significant information.

The most closely relevant articles to our work are~\cite{Rosenberg-TR11} 
and~\cite{Zhang-PLOSOne13}. Rosenberg~\cite{Rosenberg-TR11} described a three-class 
scheme for scientists' classification based on the length and thickness of the 
tail of the citation curve. Zhang in~\cite{Zhang-PLOSOne13} proposed the $h^{\prime}$ 
index as a quantitative measure to discover which scientist belongs to which one of 
those three categories.

Collectively, the present work differentiates itself by the previous studies in a 
number of factors: a) it exploits the full spectrum of information under the citation 
curve, b) it is based on the definition of new areas (not below, but above the 
citation curve), c) it penalizes those scientists with long and thin tails, d) it 
proposes an index that can be used as a filter to separate the constantly influential 
scientists from the mass producers.

\section{Penalty areas and the Perfectionism Indices}
\label{sec:areas}

In this section, we will define the {\it penalty areas} which form the basis for 
the development of the respective scientometric indices. Before proceeding further, 
we summarize in Table~\ref{symbols_tab} some symbols, their interpretations, 
and the relationships among them, which will be used throughout this article.

\begin{table}[!hbt]
\caption{Basic symbols and their interpretation.}
\label{symbols_tab}
\begin{tabular}{|c||l|c|}\hline
Symbol& Description					& Relations\\ \hline \hline
$h$   & \hi of an author				&\\
$p$   & number of articles of an author			&\\
$P$   & set of articles of an author			& $|P|=p$\\
$P_H$ & set of articles that belong in the core area	& $|P_H|=h$\\
$P_T$ & set of articles that belong in the tail area	& $|P_T|=p_T=p-h$\\
$p_T$ & number of articles that belong in $P_T$		&\\
$C$   & number of citations of an author		&\\
$C_i$ & number of citations for publication $i$		&\\
$C_H$ & number of citations for publications in $P_H$	& $C_H=\sum_{\forall i \in P_H} C_i=R^2$ \cite{Jin2007}\\
$C_T$ & number of citations for publications in $P_T$	& $C_T=\sum_{\forall i \in P_T} C_i$\\
$C_E$ & number of citations in the upper (excess) area	& $C_E=C_H-h^2$\\ \hline 
\end{tabular}

\end{table}

\subsection{The tail complement penalty area}
\label{sec:xarea}

\begin{figure}[!hbt]
\centering
\subfigure[]{\label{example_authorA}
\resizebox{5.5cm}{4.0cm}{\psfig{figure=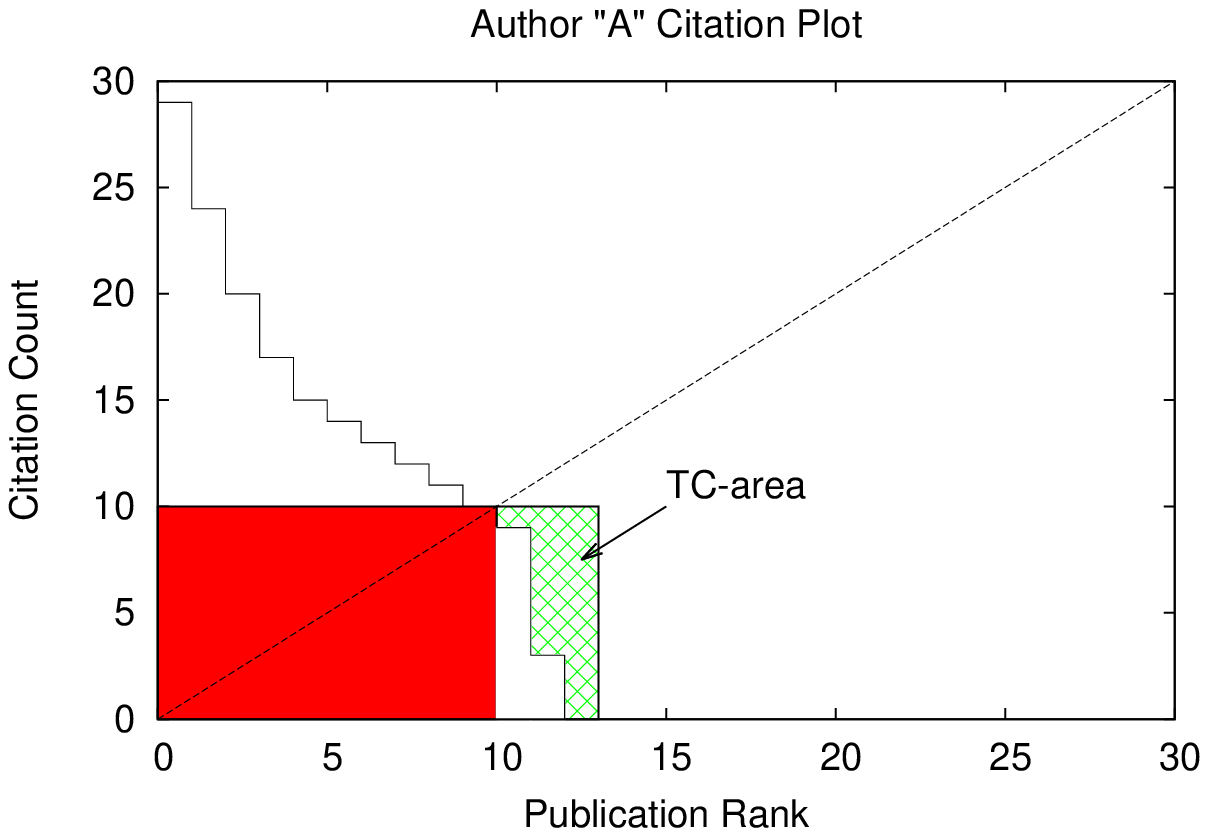,clip=}}}
\subfigure[]{\label{example_authorB}
\resizebox{5.5cm}{4.0cm}{\psfig{figure=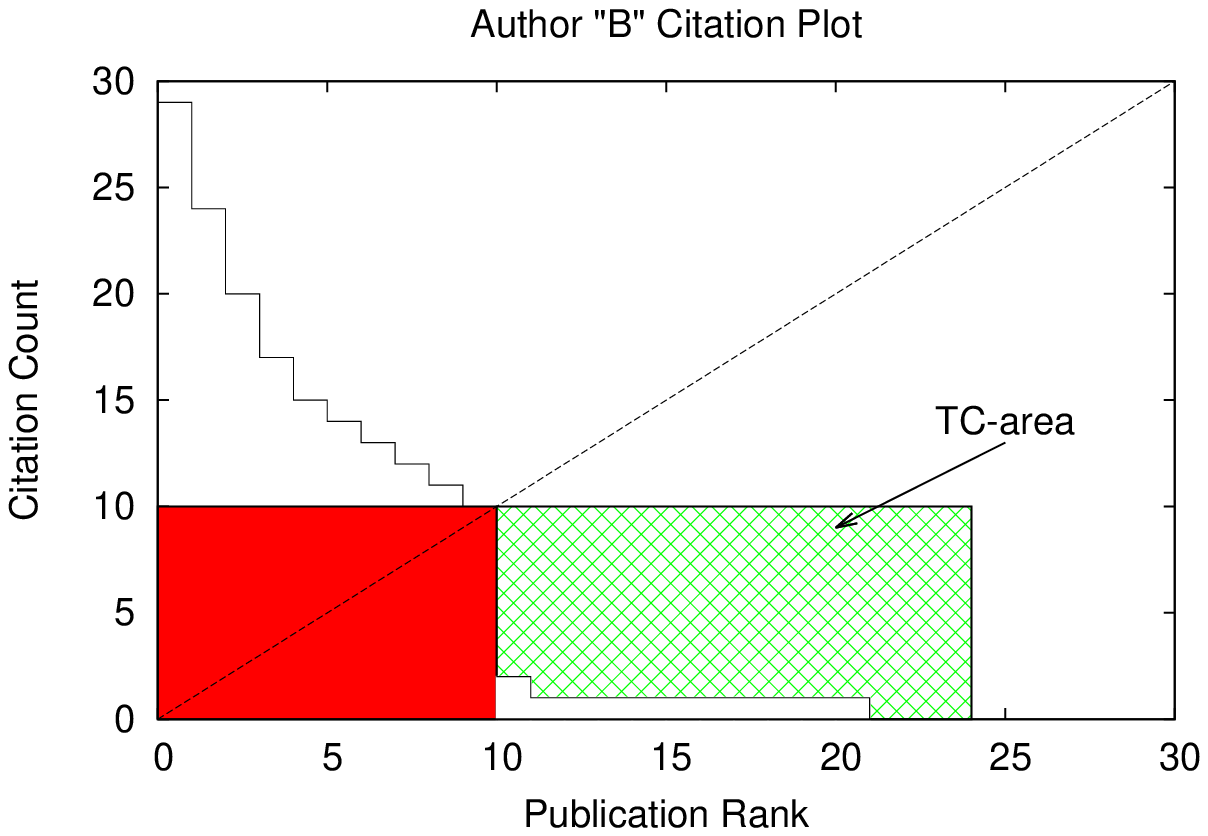,clip=}}}
\caption{Citation curves for two sample authors $A$ and $B$.}
\label{example_authorsAB}
\end{figure}

We now get back to the motivating example presented in the previous section, and 
we illustrate graphically their citation distributions (see Figure~\ref{example_authorsAB}). 
We depict with red color the h-core area of each author.

It is intuitive that long tails and light-weight tails reduce an author's articles' collective 
influence. Therefore, we argue such kind of a tail area should be considered as a ``negative" 
characteristic when assessing a scientists's performance. The closer the citations of the tail's 
articles get to the line $y=h$, the more probable it is for the scientist to increase his \hi, and 
at the same time to be able to claim that practically each and every article he publishes does 
not get unnoticed by the community.

For this purpose, we define a new area, the {\it tail complement penalty area}, denoted as 
{\it TC-area} with size $C_{TC}$. The size of the tail complement penalty area is computed as 
follows:
\begin{equation}
C_{TC} ~=~ \sum_{\forall i \in P_T}(h-C_i) ~=~ h \times (p-h) - C_T.
\label{eq_C_X}
\end{equation}

This area is depicted with the green crossing-lines pattern in Figure~\ref{example_authorsAB}, 
and fulfilling the motivation behind its definition, it is much bigger for author~$B$ than for 
author~$A$.

\subsection{The ideal complement penalty area}
\label{sec:idealarea}

\begin{figure}[!hbt]
\centerline{\psfig{figure=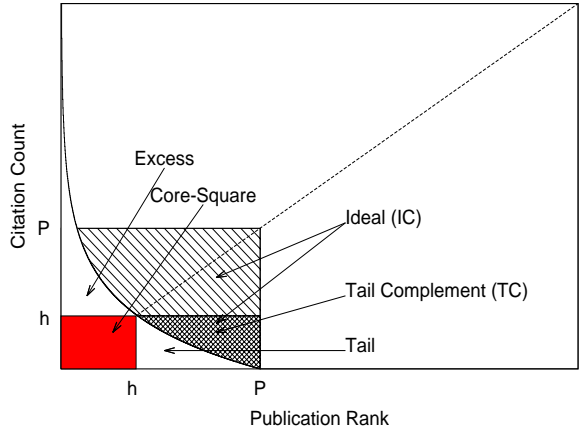,clip=,width=8cm,height=6cm}}
\caption{Graph illustrating all (existing and proposed) areas.}
\label{example_dp_fig}
\end{figure}

If we push further the idea of the tail complement penalty area, we can think that ``ideally" 
an author could publish $p$ papers with $p$ citations each and get an \hi equal to $p$. 
Thus, a square $p \times p$ could represent the minimum number of citations to achieve an \hi 
value equal to $p$. Along the spirit of penalizing long and thin tails, we can define another 
area in the citation curve: the {\it ideal complement penalty area} ({\it IC-area}), 
which is the complement of the citation curve with respect to the square $p \times p$. 
Figure~\ref{example_dp_fig} illustrates graphically the IC-area with the green crossing-lines 
pattern. The size of the IC-area ($C_{IC}$) can be computed as follows: 
\begin{equation}
C_{IC} ~=~ \sum_{\forall i \in P\ \wedge\ C_i<p}{(p-C_i)}.
\label{eq_CI}
\end{equation}

Apparently, this area does not depend on the \hi value, as it holds for the case of the TC-area. 
Notice that the IC-area includes the TC-area defined in the previous paragraph. We realize of 
course that it is hard (if possible at all) to find scientists -- with sufficiently large \hi, 
and -- with zero-sized TC-area, since citation distributions approximately follow power 
laws~\cite{Burrell13,Clauset-SIAM-Review09,Newman-SIAM-Review03}. Therefore, the index derived by this 
area is not expected to provide significant insights into scientists' performance.

\subsection{The new scientometric perfectionism indices: $PI$ and $XPI$}
\label{sec:PI}

The definition of the penalty areas in the previous subsection, allows us to design two 
new metrics which will act as the filter to separate influential from mass producers. 
Firstly, let us introduce the concept of {\it Parameterized Count}, $PC$, as follows: 
\begin{equation}
\label{eq-pc}
PC ~=~ \kappa*h^2 + \lambda*C_E + \mu*C_T 
\end{equation}
where $\kappa,\lambda,\mu$ are integer values. Apparently:
\begin{itemize}
\item when $\kappa=\lambda=\mu=1$, then it holds that $PC=C$, 
\item when $\kappa=1\ \wedge\ \lambda=\mu=0$, then $PC=h^2$, 
\item when $\lambda=1\ \wedge\ \kappa=\mu=0$, then $PC=C_E=e^2=C_h-h^2$, 
\item when $\mu=1\ \wedge\ \kappa=\lambda=0$, then $PC=C_T$.
\end{itemize}
By assigning positive values to $\kappa$ and $\lambda$, but negative values to $\mu$, 
we can favor authors with short and thick tails in the citation curve. Even in this way, 
we cannot differentiate between the authors~$A$ and~$B$ of our example.

For this reason, instead of using the tail of the citation curve, we use the tail complement 
penalty area. Thus, similarly to Equation~\ref{eq-pc}, we define the concept of 
{\it Perfectionism Index based on TC-area} as follows:
\begin{equation}
PI ~=~ \kappa*h^2 + \lambda*C_E - \nu*C_{TC} 
\label{eq_PI}
\end{equation}

In the experiments that will be reported in the next sections, we will use the values of 
$\kappa=\lambda=\nu=1$. These default values give a straightforward geometrical notion of 
the newly defined metric. Noticeably, it will appear that $PI$ can get negative values. Thus:
\begin{itemize}
\item if an author has $PI < 0$, then we characterize him as a {\it mass producer}, 
\item if an author has $PI > 0$, then we characterize him as an {\it influential}.
\end{itemize}

In the same way as the $PI$'s definition, we define an extremely perfectionism metric, the {\it Extreme Perfectionism Index},
taking into account the ideal complement penalty area, as follows:
\begin{equation}
XPI = \kappa*h^2 + \lambda*C_E + \mu*C_T - \nu*C_{IC}.
\label{eq_XPI}
\end{equation}

As in the previous case, we will assume that $\kappa=\lambda=\mu=\nu=1$. We will show 
in the experiments, that very few authors have positive values for this metric. Using the 
previously defined perfectionism indices, the resulting values for authors~$A$ and~$B$ are shown in 
Table~\ref{tab_exampleAB_authors_results}. Author~$A$ has greater values than author~ $B$ for 
both~$XPI$ and~$PI$ perfectionism indices. This is a desired result.

\begin{table}[!hbt]
\caption{Traditional and proposed indices for authors~$A$ and~$B$.}
\label{tab_exampleAB_authors_results}
\begin{center}
\begin{tabular}{|c||cccccccccc|} \hline
Author & $p$& $C$ & $h$& $C_T$& $C_E$& $C_H$& $C_{TC}$& $PI$& $C_{IC}$& $XPI$ \\ \hline\hline
$A$ & 13 & 177 & 10 & 12 & 65 & 165 & 18 & 147& 33  & 144 \\
$B$ & 24 & 177 & 10 & 12 & 65 & 165 & 128& 37 & 404 & -227\\ \hline
\end{tabular}

\end{center}
\end{table}

\begin{figure}[!tb]
\centering
\subfigure[]{\label{h10_example4} \resizebox{5.7cm}{4.7cm}{\psfig{figure=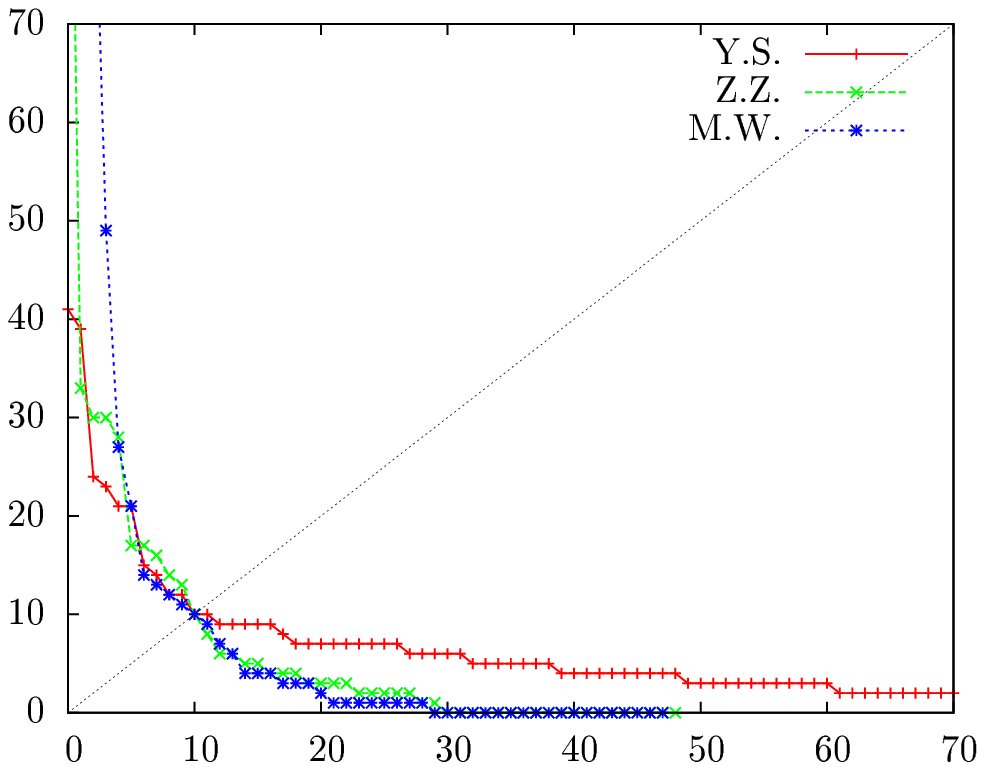,clip=}}}
\subfigure[]{\label{h10-15_example7} \resizebox{5.7cm}{4.7cm}{\psfig{figure=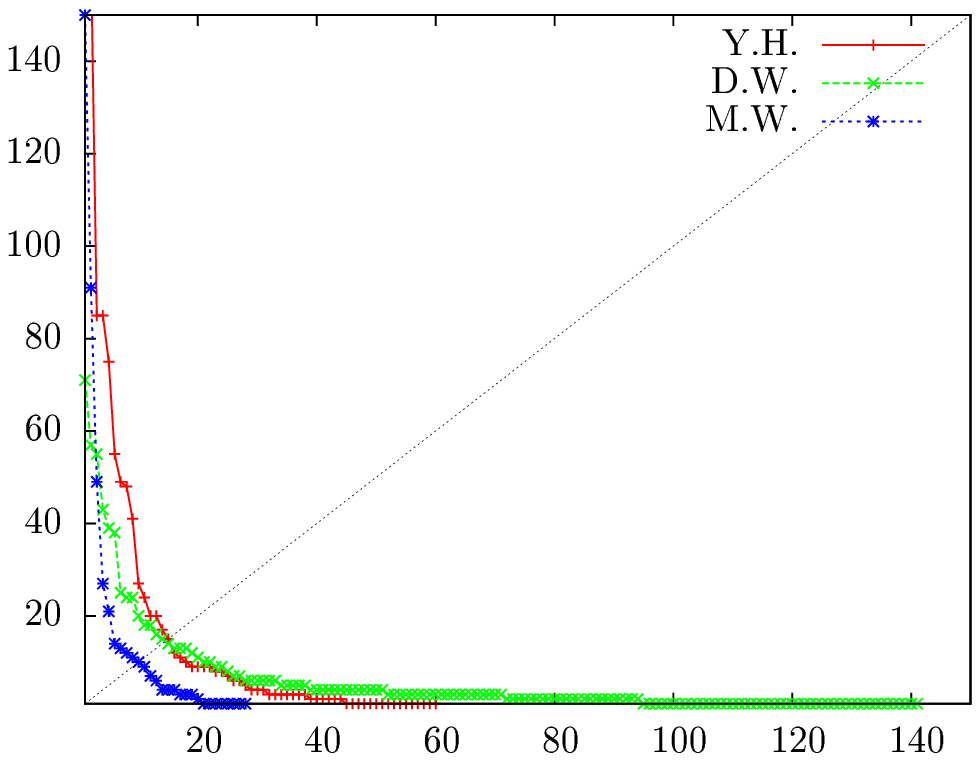,clip=}}}
\subfigure[Adjusted form of Figure~\ref{real_examples_1}\subref{h10-15_example7}.]{\label{h10-15_example7_scaled}\resizebox{6cm}{5.2cm}{\psfig{figure=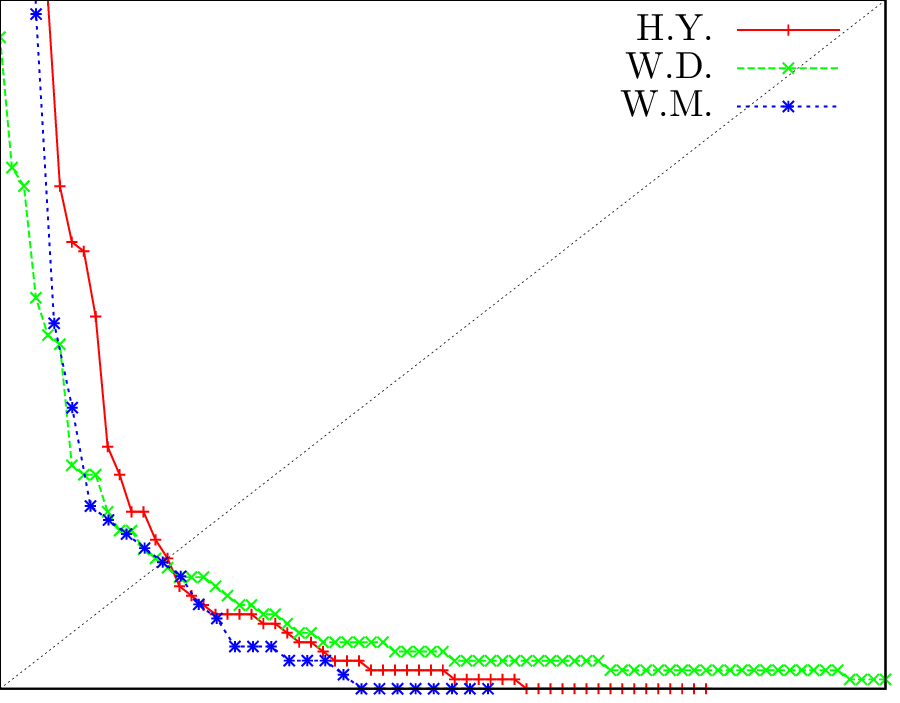,clip=}}}
\caption{Real examples.}
\label{real_examples_1}
\end{figure}

Before proceeding to the next section, which describes the detailed experiments that demonstrate the 
merits of the new indices, we provide an additional example of five authors with different 
publishing patterns, as an extension to our artificial motivating example which was presented 
in the beginning of the article. 
We use only initials but they refer to real persons and their data.
In Table~\ref{tab_real_examples} we present the raw data 
(i.e., \hi, number of publications $p$ and number of citations $C$) of 
5~authors\footnote{We selected authors with relatively small number of publications and 
citations for better readability of the figures.} selected from Microsoft Academic 
Search\footnote{\url{http://academic.research.microsoft.com/}}. The last column 
shows the calculated $PI$ values, which can be positive as well as negative numbers. 
In Figure~\ref{real_examples_1} we present citation plots for these five authors.

In Figure~\ref{h10_example4} we compare three authors: S.Y., Z.Z. and W.M. 
They all have an \hi equal to~10. Note that S.Y. has a comparatively large number of publications 
but the citation curve is cropped to focus on the lower 
values. As he has the bigger and longest tail (red line), he could be characterized as a ``mass producer". 
This is reflected in a $PI$ value of $-2505$ as shown in 
Table~\ref{tab_real_examples}. Z.Z. (green line) has shorter 
tail than S.Y. and higher excess area. From the same table we remark that his $PI$ value 
is 11 (i.e. close to zero). Finally, the last author of the example, W.M. (blue line), 
has similar tail with Z.Z. but a bigger excess area ($e^2$). Definitely, 
he demonstrates the ``best" citation curve out of the three authors of the example. 
In fact, his $PI$ score is 717, higher than the respective figure of the other two authors.

\begin{table}[!tb]
\caption{Computed \hi and PI values for 5 sample authors.}
\label{tab_real_examples}
\begin{center}
\begin{tabular}{|l@{}||r|r|r|r|}\hline
\multirow{2}{*}{\bf Author } & \multicolumn{1}{|c|}{\multirow{2}{*}{\bf $h$}}& \multicolumn{1}{|c|}{\multirow{2}{*}{\bf $p$}}& \multicolumn{1}{|c|}{\multirow{2}{*}{\bf $C$}}& \multicolumn{1}{|c|}{\multirow{2}{*}{\bf $PI$}}\\
              &    &    &     & \\\hline\hline
H. Y.& 15 & 105& 2040& 690\\ 
W. D.& 15 & 259& 1137& -2523\\ 
W. M.   & 10 & 48 & 1097& 717\\ 
S. Y.      & 10 & 319& 585 & -2505\\ 
Z. Z.   & 10 & 49 & 391 & 11\\ \hline
\end{tabular}

\end{center}
\end{table}

In Figure~\ref{h10-15_example7}, again we compare three authors: H.Y., W.D. 
and W.M. The first two have \hi value equal to 15. Comparing those, it seems 
that H.Y. (red line) has a better citation curve than W.D. (green line) 
because he has a shorter tail and a bigger excess area. Indeed, the first one has $PI=717$, 
whereas the second one has $PI=-2523$. W.M. (blue line) has a smaller tail and a 
big excess area but since there is a difference in \hi we cannot say for sure if he must be 
ranked higher or not than the others.

In Figure~\ref{h10-15_example7_scaled} we have scaled the citation plots so that all lines  
cut the line $y=x$ at the same point. From this plot, it is shown that W.M. has a better 
curve than W.D. because he has a shorter tail and a bigger excess area. When comparing 
W.M. to H.Y., we see that the latter has a longer tail but also a bigger 
excess area. Both curves show almost the same symmetry around the line $y=x$. That is why they 
both have similar $PI$ values. This is a further positive outcome as authors with different 
quantitative characteristics (say, a senior and a junior one) may have similar qualitative 
characteristics, and thus be classified together.

\section{Experiments}
\label{sec:experiments}

In this section, we will present the results of the evaluation of the proposed indexes. The 
primary goal of our experimentation is to study the merits of the $PI$ index, since our results 
confirmed that the severe penalty that $XPI$ imposes makes it a less useful scientometric tool. 
Firstly, we will explain the procedures for dataset acquisition, then we will present their 
characteristics, and finally we will give the results that concern the evaluation of~$PI$.

\subsection{Datasets acquisition and characterization}
\label{subsec:dataset}

During the period December~2012 to April~2013, we compiled 3~datasets. The first one 
consists of randomly selected authors (named ``Random" henceforth). The second one includes highly 
productive authors (named ``Productive"). The last one consists of authors in the top \hi list 
(named ``Top h"). The publication and the citation data were extracted from the Microsoft 
Academic Search (MAS) database using the MAS API. 

The dataset ``Random" was generated as follows: We fetched a list of about~100000 authors 
belonging to the ``Computer Science" domain as tagged by MAS. MAS assigns at least three 
sub-domains to every author. These three sub-domains may not all belong to the same domain 
(e.g., Computer Science). For example, an author may have two sub-domains from Computer 
Science and one from Medicine. We kept only the authors who have their first three sub-domains 
belonging to the domain of Computer Science. From this set, we randomly selected 500~authors 
with at least 10~publications and at least 1~citation.

The dataset ``Productive" was generated as follows: from the set of 100000~Computer Science 
authors we selected the top-500 most productive. The least productive author from this sample 
has 354~publications.

The third dataset named ``Top h" was generated by querying the MAS Database for the top-500 
authors in the ``Computer Science" domain ordered by \hi.

Table~\ref{table_stats} summarizes the information about our datasets with respect to the number 
of authors (line: \# of authors), number of publications (line: \# of publications), number 
of citations (line: \# of Citations) and average/min/max numbers of citations and publications 
per author.

\begin{table}[!hbt]
\caption{Statistics of the datasets used in our study.}
\label{table_stats}
\begin{center}
\begin{tabular}{|r|ccc|}\hline
                   & \bf{Random} & \bf{Productive} & \bf{Top h}\\\hline\hline
\# of authors      & 500         & 500             & 500 \\
\# of publications & 25679       & 223232          & 149462 \\
\#P/Author         & 51          & 446             & 298 \\
Min \#P/Author     & 10          & 354             & 92 \\
Max \#P/Author     & 768         & 1172            & 1172 \\
\# of Citations    & 410280      & 3197880         & 5015971 \\
\# Cit/Author      & 820         & 6395            & 10031 \\
Min \#Cit/Author   & 1           & 25              & 4405 \\
Max \#Cit/Author   & 47263       & 47263           & 47263\\ \hline
\end{tabular}

\end{center}
\end{table}

\begin{figure}[!htb]
\centering
\subfigure[$h$-index]{\label{y0_c0_s1_distr_hindex_} \resizebox{5.5cm}{3.8cm}{ \psfig{figure=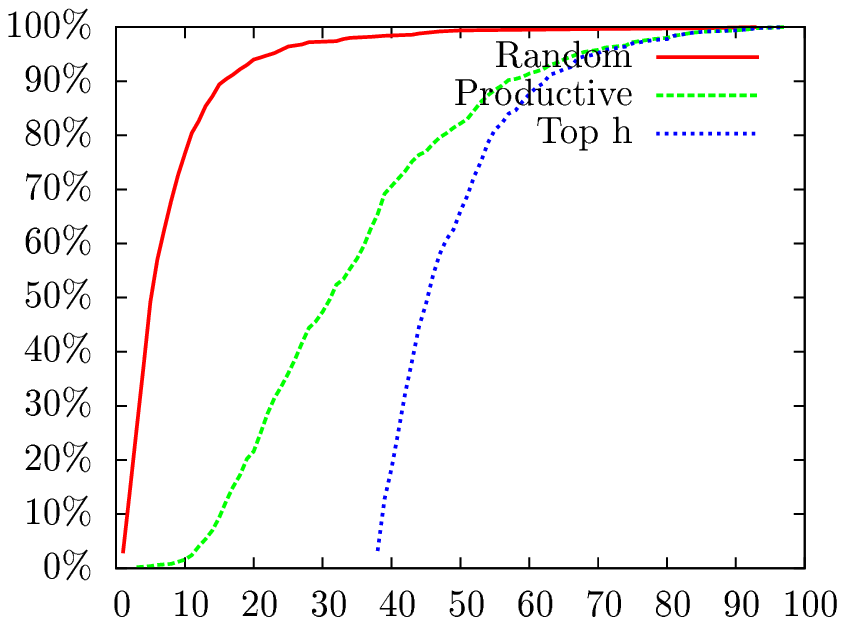,clip=}}}
\subfigure[$h$-index]{\label{y0_c0_s1_distr_hindex__grouped} \resizebox{5.5cm}{3.8cm}{ \psfig{figure=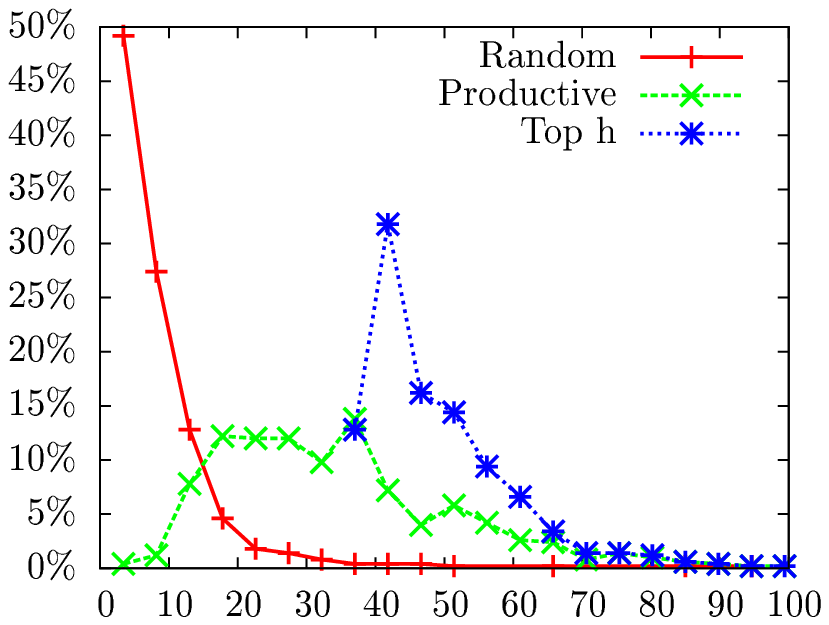,clip=}}}
\subfigure[$C$ (* 10000)]{\label{y0_c0_s1_distr_C_} \resizebox{5.5cm}{3.8cm}{ \psfig{figure=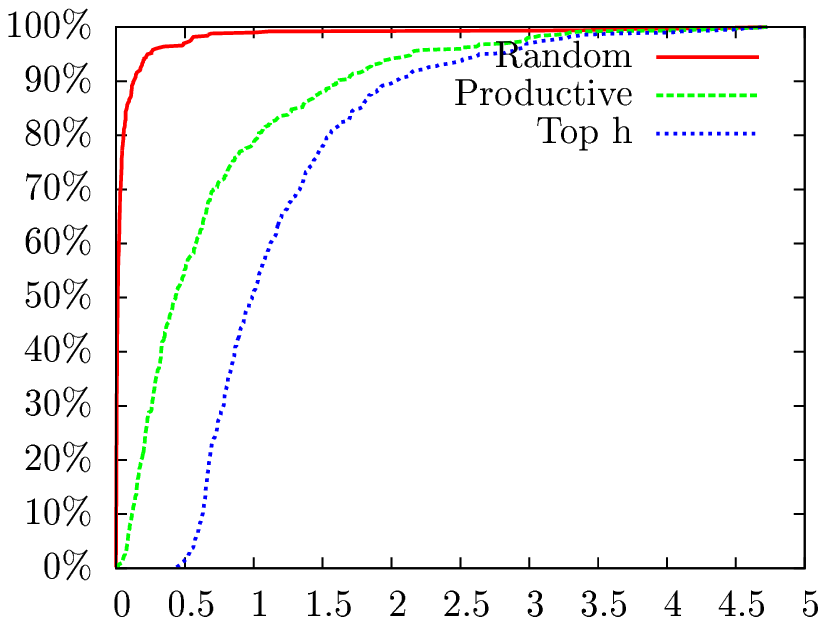,clip=}}}
\subfigure[$C$ (* 10000)]{\label{y0_c0_s1_distr_C__grouped} \resizebox{5.5cm}{3.8cm}{ \psfig{figure=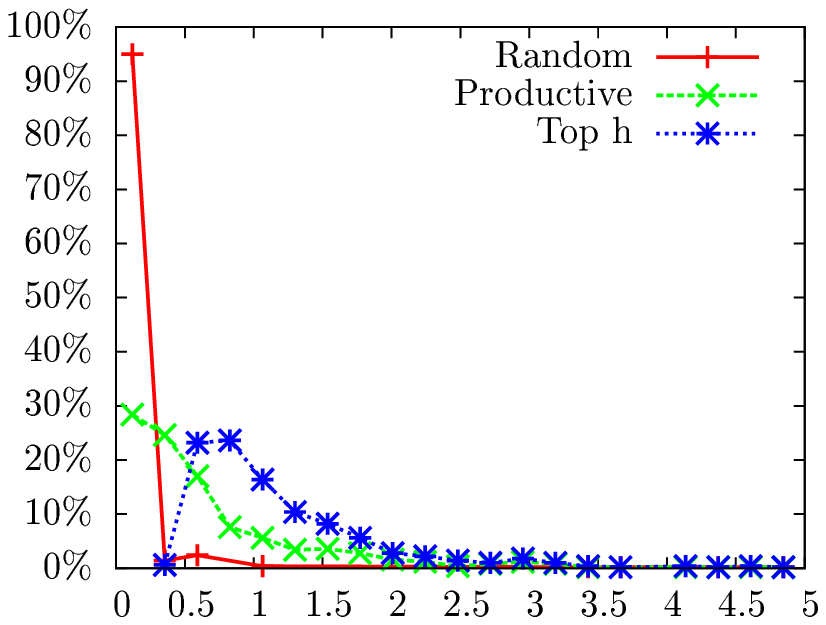,clip=}}}
\subfigure[$m$]{\label{y0_c0_s1_distr_m_} \resizebox{5.5cm}{3.8cm}{ \psfig{figure=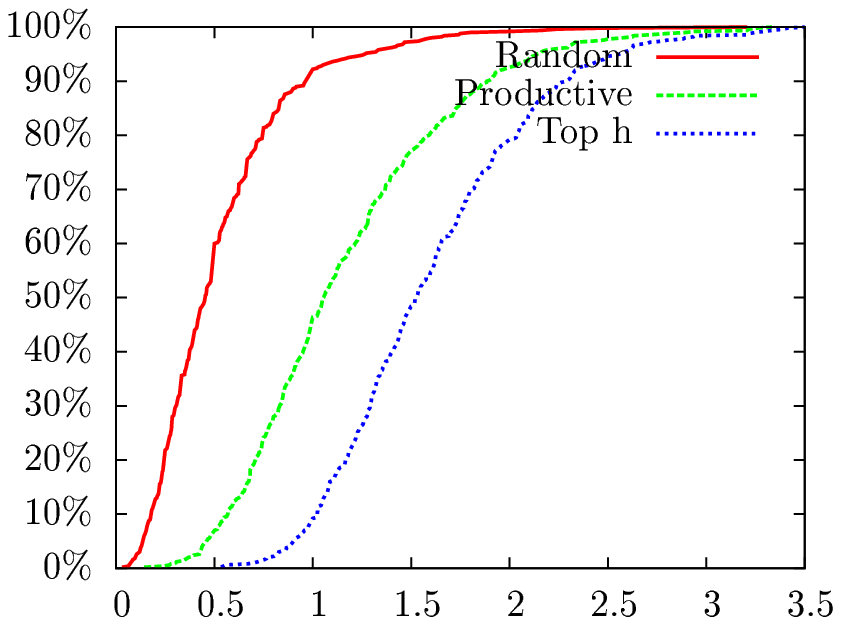,clip=}}}
\subfigure[$m$]{\label{y0_c0_s1_distr_m__grouped} \resizebox{5.5cm}{3.8cm}{ \psfig{figure=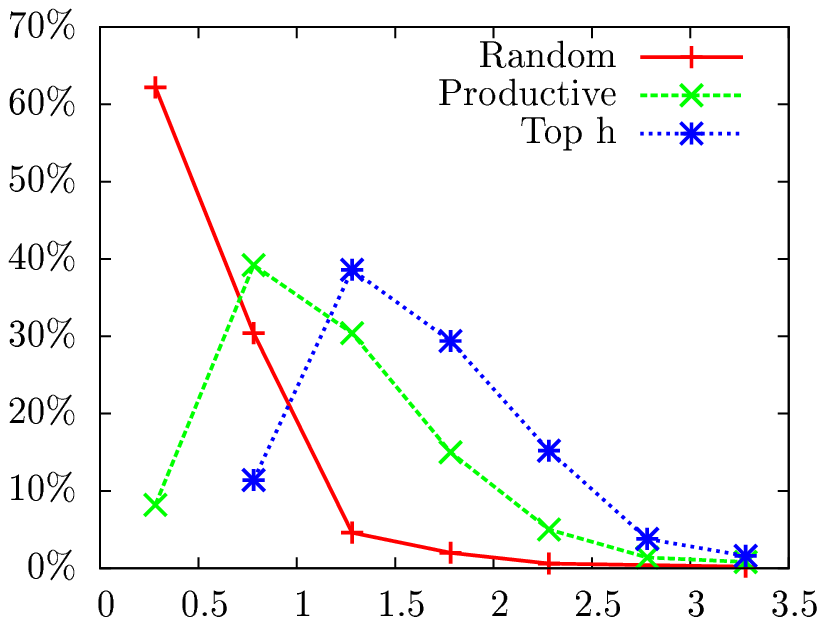,clip=}}}
\subfigure[$p$ (* 1000)]{\label{y0_c0_s1_distr_p_} \resizebox{5.5cm}{3.8cm}{ \psfig{figure=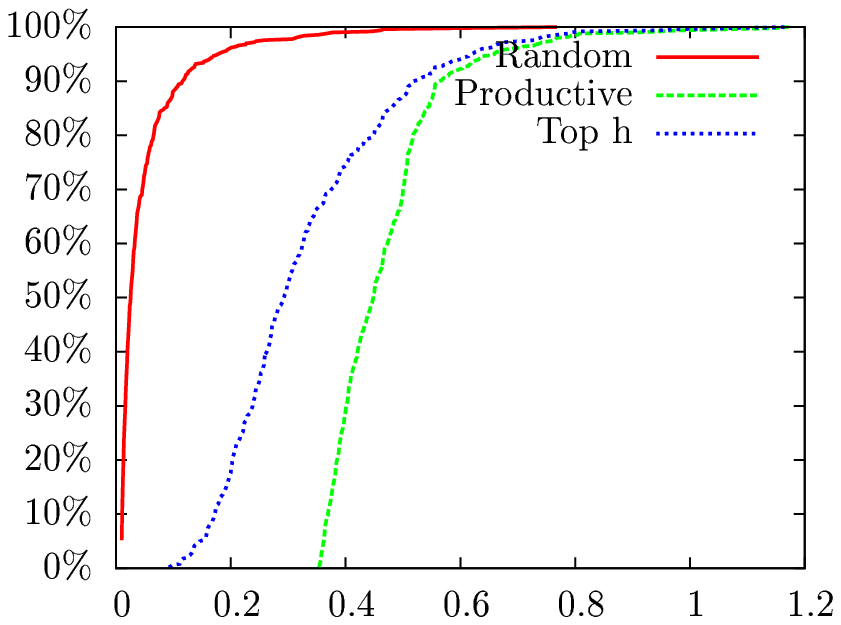,clip=}}}
\subfigure[$p$ (* 1000)]{\label{y0_c0_s1_distr_p__grouped} \resizebox{5.5cm}{3.8cm}{ \psfig{figure=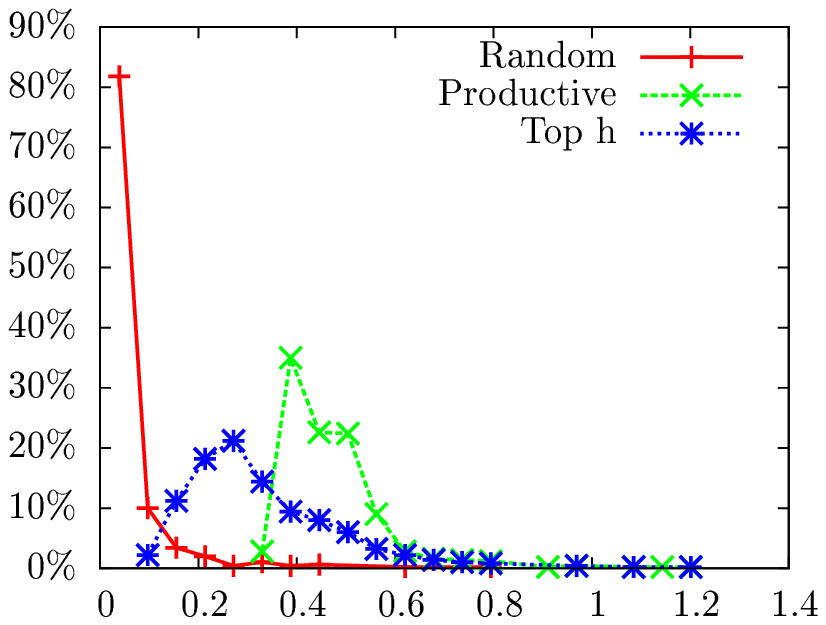,clip=}}}
\caption{Distributions of \hi, $m$, $p$, $C$ indices (Left plots: CDFs. Right plots: PDFs)}
\label{fig_plot_all_plot_basic}
\end{figure}

Figure~\ref{fig_plot_all_plot_basic} shows the distributions for the values of \hi, $m$, $C$ 
and $p$. The $m$~index was defined in Hirsch's original article~\cite{Hirsch-PNAS05} and 
is explained (quoting Hirsch's text) in the next paragraph. Plots are illustrated in pairs. 
The ones on the left show cumulative distributions. For example, in Figure~\ref{y0_c0_s1_distr_hindex_} 
we see that 80\% of the authors in the sample ``Random" (red line) have \hi less than~10. 
It is obvious that the sample ``Top h" (blue line) has higher values for the \hi. 
Figures~\ref{y0_c0_s1_distr_C_} and~\ref{y0_c0_s1_distr_C__grouped} show the distributions 
for the total number of citations. As expected the sample ``Top h" has the highest values. 

Figures~\ref{y0_c0_s1_distr_m_} and~\ref{y0_c0_s1_distr_m__grouped} show the distributions for 
the $m$ value. Recall its definition from \cite{Hirsch-PNAS05}: A value of $m\approx 1$ (i.e., 
an \hi of 20 after 20 years of scientific activity), characterizes a successful scientist. A 
value of $m \approx 2$ (i.e., an \hi of 40 after 20 years of scientific activity), characterizes 
outstanding scientists, likely to be found only at the top universities or major research 
laboratories. A value of $m \approx 3$ or higher (i.e., an \hi of 60 after 20 years, or 90 after 
30 years), characterizes truly unique individuals. Indeed, only a few authors have $m>3$.

Figures~\ref{y0_c0_s1_distr_p_} and~\ref{y0_c0_s1_distr_p__grouped} illustrate the distributions 
for the total number of publications. It is obvious that in the ``Random" sample (red line) 
there are relatively low values for the total number of publications. Also, as expected, 
the distribution for the ``Productive" sample has the highest values for the total number of 
publications.

We have conducted further experiments to study the behavior of other indices such as 
$\alpha$~\cite{Hirsch-PNAS05} and $e^2$~\cite{Zhang-PLOSOne09}. However, the results did not 
carry any significant information, and therefore, the figures for these factors are not presented.

\subsection{Does $PI$ offer new insights about the impact and publication habits of scientists?}
\label{subsec:need}

The first question that needs to be answered is whether a new index offers something new and 
different compared to the existing (hundreds of) indices. The answer is positive; our metric 
separates the rank tables into two parts independently from the rank positions.

\begin{figure}[!bt]
\centering
\subfigure[$h$ vs. $C$ (unioned)]{\label{qqplots_hindex_c0_s1_y0_vs_C_c0_s1_y0_union} \resizebox{5.5cm}{3.9cm}{ \psfig{figure=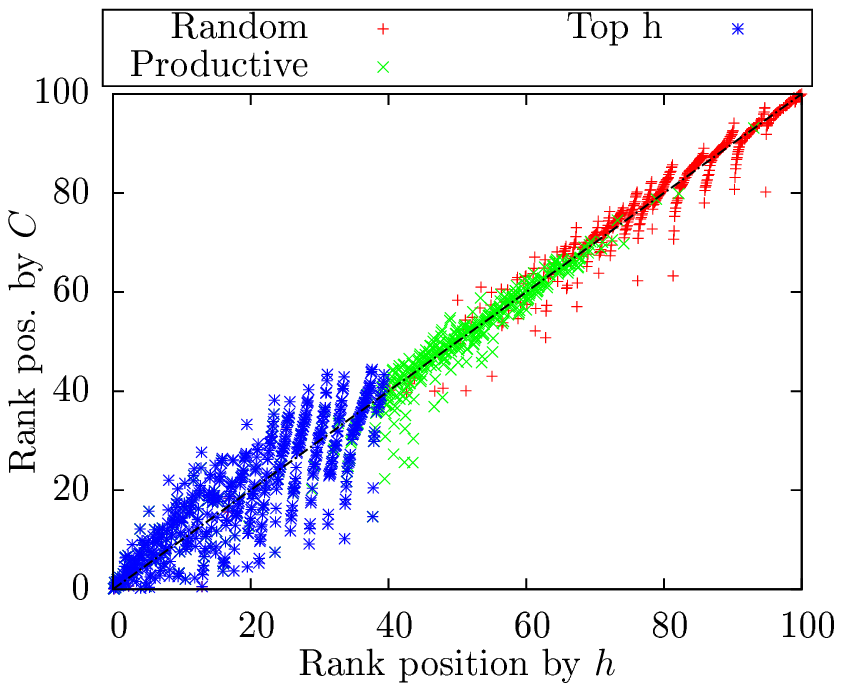,clip=}}}
\subfigure[$h$ vs. $PI$ (unioned)]{\label{qqplots_hindex_c0_s1_y0_vs_PI_c0_s1_y0_union} \resizebox{5.5cm}{3.9cm}{ \psfig{figure=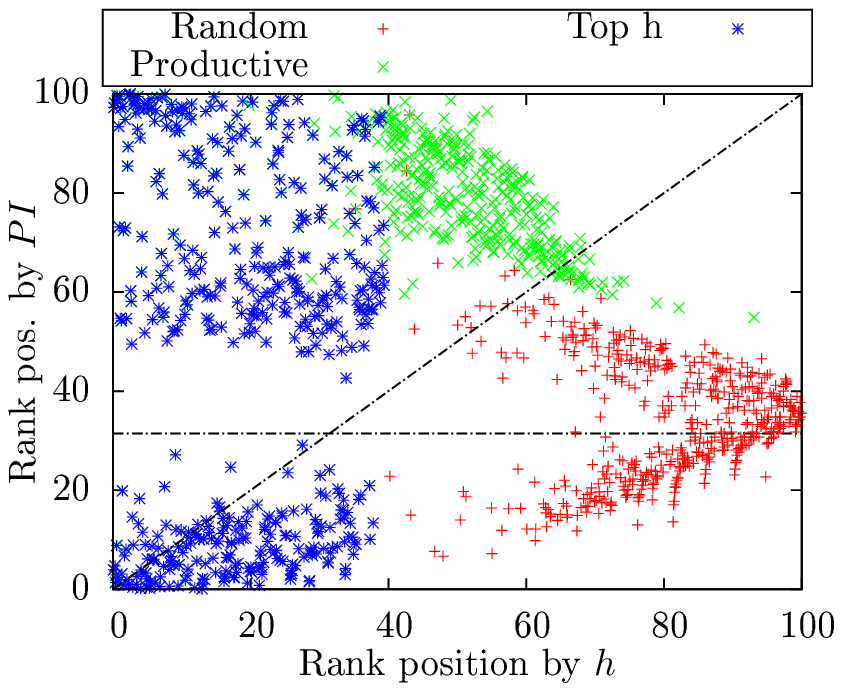,clip=}}}
\subfigure[$C$ vs. $PI$ (unioned)]{\label{qqplots_C_c0_s1_y0_vs_PI_c0_s1_y0_union} \resizebox{5.5cm}{3.9cm}{ \psfig{figure=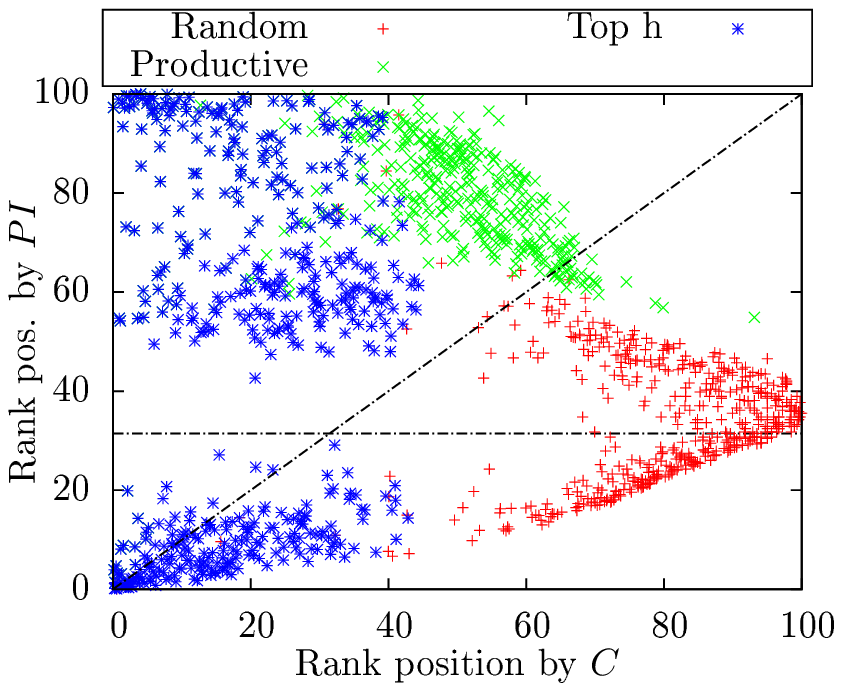,clip=}}}
\subfigure[$C/p$ vs. $PI$ (unioned)]{\label{qqplots_citp_c0_s1_y0_vs_PI_c0_s1_y0_union} \resizebox{5.5cm}{3.8cm}{ \psfig{figure=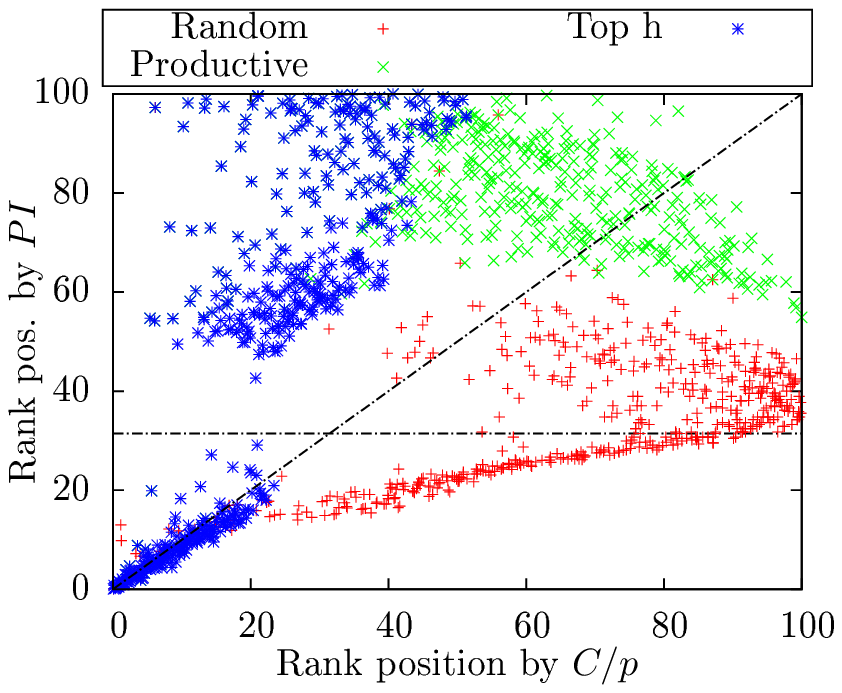,clip=}}}
\subfigure[$C/p$ vs. $h$ (unioned)]{\label{qqplots_citp_c0_s1_y0_vs_hindex_c0_s1_y0_union} \resizebox{5.5cm}{3.8cm}{ \psfig{figure=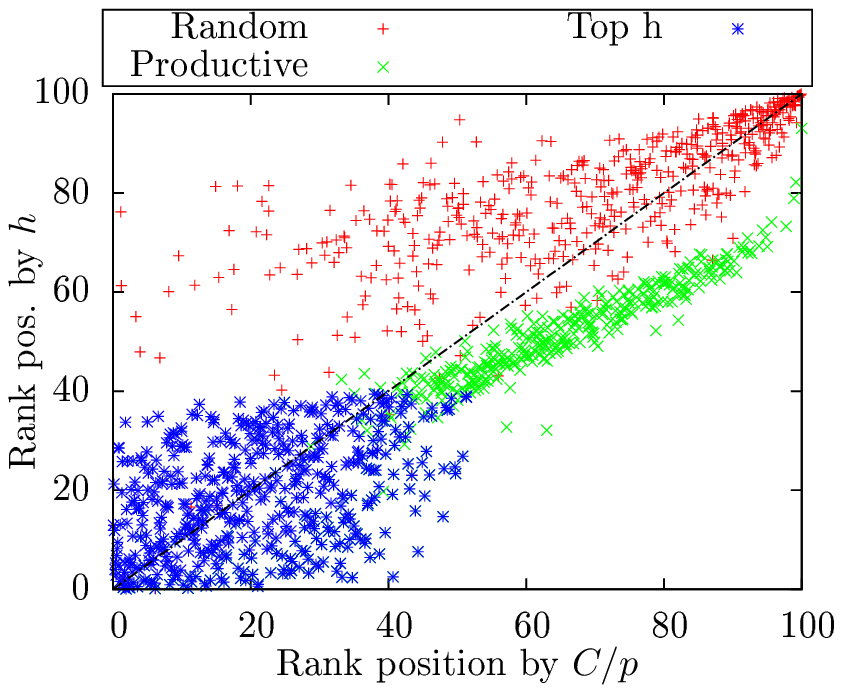,clip=}}}
\caption{Correlation of PI to standard bibliometric indices. (Q-Q plots: X- and Y-axis denote normalized rank positions (\%).)}
\label{plot_all_qqplot_1}
\end{figure}

In Figure~\ref{qqplots_hindex_c0_s1_y0_vs_C_c0_s1_y0_union} the x-axis denotes the rank 
position (normalized percentagewise) of an author by \hi, whereas the y-axis denotes the 
rank position by the total number of citations ($C$). Each point denotes the position of 
an author ranked by the two metrics. Note that all three samples are merged but if the 
point is blue, then the author belongs to the ``Top h" sample, if the point is green then 
he belongs to the ``Productive" sample etc. If an author belongs to more than one sample, 
then only one color is visible since the bullet overwrites the previous one. From 
Figure~\ref{qqplots_hindex_c0_s1_y0_vs_C_c0_s1_y0_union} the outcomes are:
\begin{itemize}
\item 
``Top h" authors are ranked in the first 40\% of the rank table by \hi, as well as in the top 
40\% by the total number of citations ($C$).
\item 
``Productive" authors are mainly ranked by \hi between 30\% and 70\%. The rank positions by $C$ 
are between 20\% and 70\%.
\item 
``Random" authors are mainly ranked below 60\% for both metrics with some outliers in the range 
0-60\%, mostly by $C$.
\end{itemize}
The aforementioned conclusions are as expected; it also occurs that \hi ranking does not 
differ significantly from the $C$ ranking; i.e., they are correlated which is 
consistent\footnote{\url{http://michaelnielsen.org/blog/why-the-h-index-is-virtually-no-use/}} 
with earlier findings~\cite{Spruit-TR12}.

In Figure~\ref{qqplots_hindex_c0_s1_y0_vs_PI_c0_s1_y0_union} the \hi ranking is compared to 
$PI$ ranking. It can be seen that there is no correlation between $PI$ and \hi. Note that the 
horizontal line at about 32\% (also shown later in Table~\ref{positive_tab}) shows the cut 
point of $PI$ for the zero value. Authors that reside below this line have $PI>0$ and authors 
above this line have $PI<0$.
\begin{itemize}
\item 
``Top h" authors are split to two groups. The first group is ranked in the top 20\% of the $PI$ 
rank table. The second group is ranked in the last 50\%. These two groups are also separated by 
the zero line of $PI$.
\item 
``Productive" authors are almost all ranked at lower positions by $PI$ than by \hi. Almost all 
points reside above the $PI$ zero line and also above the line $y=x$ (with some exceptions at 
about 65-70\% of the rank list).
\item 
``Random" authors are also generally higher ranked by $PI$ than by \hi. They are also split into 
two groups by the line $PI=0$. 
\end{itemize}
From the above, it seems that $PI$ is not correlated to \hi, whereas the line $PI=0$ plays the 
role of a symmetric axis. Thus, it emerges as the key value that separates the ``influential" 
authors from the ``mass producers".

In Figure~\ref{qqplots_C_c0_s1_y0_vs_PI_c0_s1_y0_union} we compare $PI$ ranking against 
$C$ (total number of citations) ranking. It is expected that the plot would be similar to 
Figure~\ref{qqplots_hindex_c0_s1_y0_vs_PI_c0_s1_y0_union} based on the similarity of \hi with~$C$.

In Figures~\ref{qqplots_citp_c0_s1_y0_vs_PI_c0_s1_y0_union} 
and~\ref{qqplots_citp_c0_s1_y0_vs_hindex_c0_s1_y0_union} we compare \hi and $PI$ with the 
average number of citations per publication ($C/p$) ranking. It is apparent that $PI$ is not 
correlated to $C/p$. \hi is also uncorrelated to $C/p$, however the points of the qq-plot in 
Figure~\ref{qqplots_citp_c0_s1_y0_vs_hindex_c0_s1_y0_union} are closer to the line $x=y$ than 
the points of Figure~\ref{qqplots_citp_c0_s1_y0_vs_PI_c0_s1_y0_union}.

Conclusively, the $PI$ ranking is not correlated to \hi, $C$ and $C/p$.

\subsection{Aggregate analysis of the datasets}
\label{subsec:stats}

\begin{figure}[!hbt]
\centering
\subfigure[$C_T$ (* 1000)]{\label{y0_c0_s1_distr_C_T_} \resizebox{5.5cm}{3.1cm}{ \psfig{figure=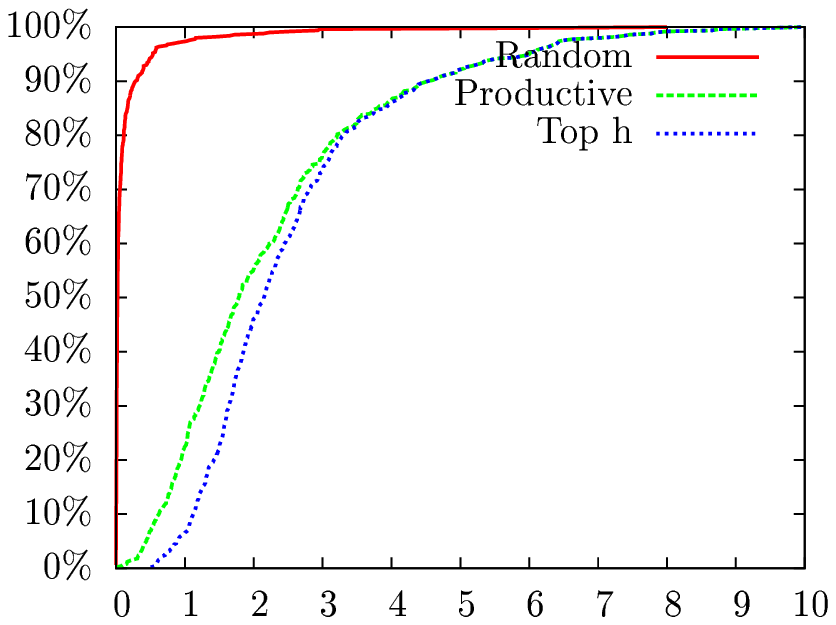,clip=}}}
\subfigure[$C_T$ (* 1000)]{\label{y0_c0_s1_distr_C_T__grouped} \resizebox{5.5cm}{3.1cm}{ \psfig{figure=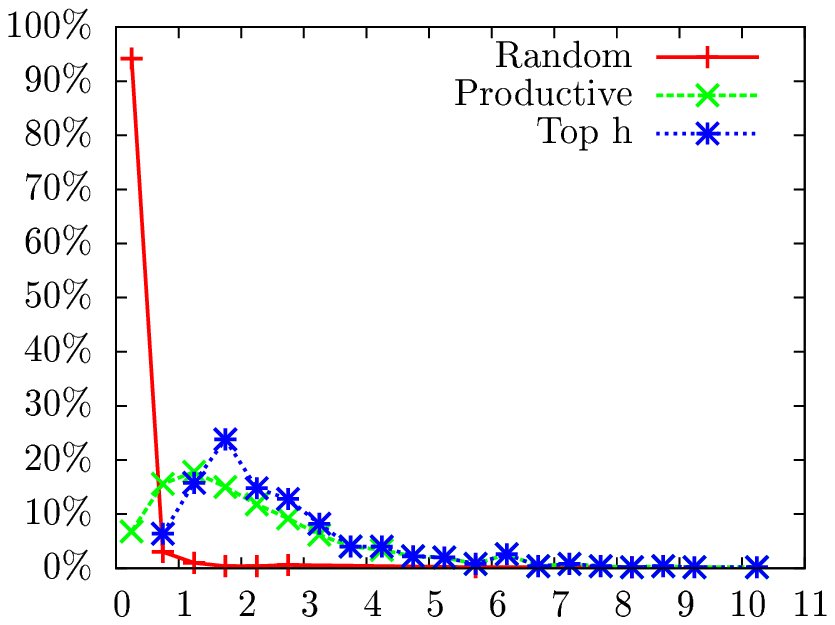,clip=}}}
\subfigure[$C_{TC}$ (* 10000)]{\label{y0_c0_s1_distr_C_X_} \resizebox{5.5cm}{3.1cm}{ \psfig{figure=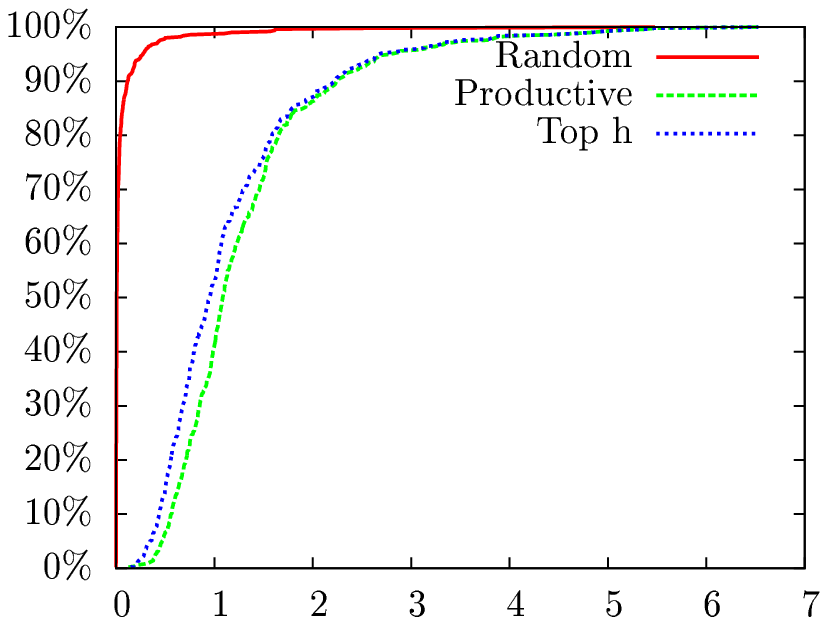,clip=}}}
\subfigure[$C_{TC}$ (* 10000)]{\label{y0_c0_s1_distr_C_X__grouped} \resizebox{5.5cm}{3.1cm}{ \psfig{figure=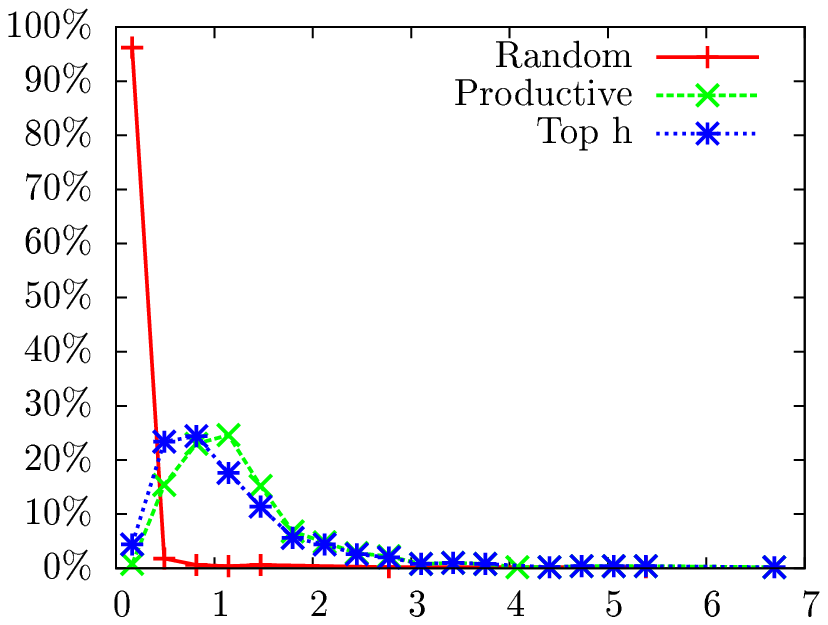,clip=}}}
\subfigure[$C_{IC}$ (* 100000)]{\label{y0_c0_s1_distr_C_I_} \resizebox{5.5cm}{3.1cm}{ \psfig{figure=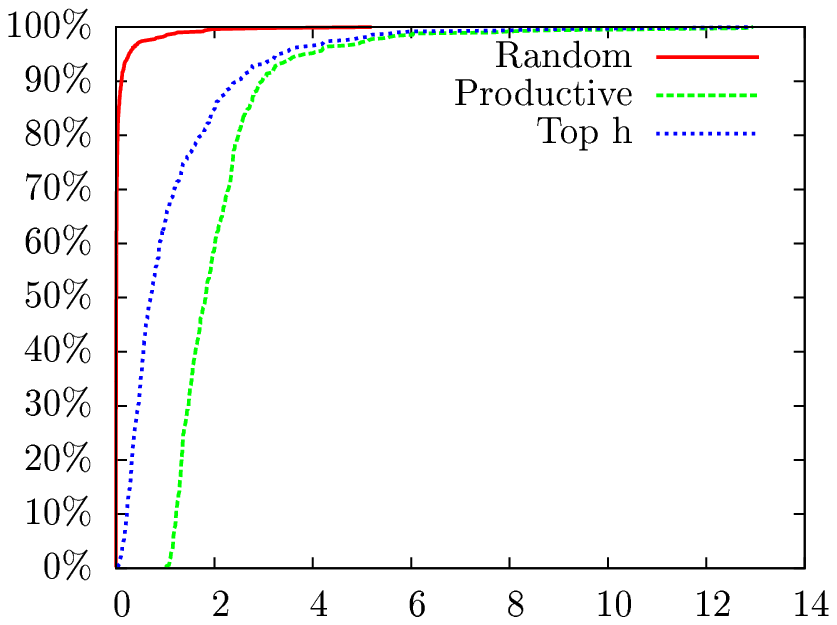,clip=}}}
\subfigure[$C_{IC}$ (* 100000)]{\label{y0_c0_s1_distr_C_I__grouped} \resizebox{5.5cm}{3.1cm}{ \psfig{figure=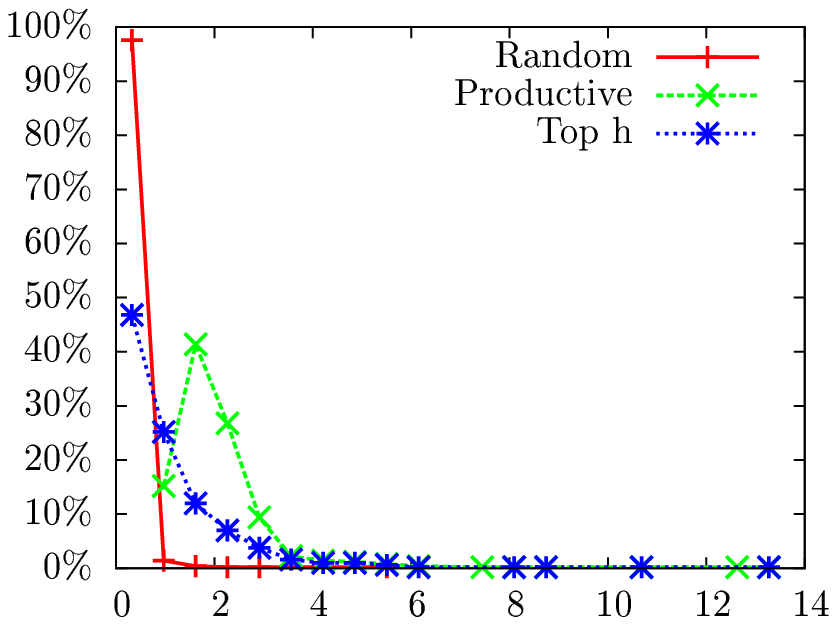,clip=}}}
\caption{Distributions of $C_T$, $C_{TC}$ (tail complement), $C_{IC}$ (ideal complement). (Left plots: CDFs. Right plots: PDFs) }
\label{fig_plot_all_plot_our_areas2}
\end{figure}

Figure~\ref{fig_plot_all_plot_our_areas2} shows the distributions for the areas defined in the 
previous section. In particular, Figures~\ref{y0_c0_s1_distr_C_T_} and~\ref{y0_c0_s1_distr_C_T__grouped} 
illustrate the distributions for the $C_T$ (tail) area. It seems that the ``Top h" cumulative 
distribution is very similar to the ``Productive" one, however, the ``Top h" distribution has 
slightly higher values.

Figures~\ref{y0_c0_s1_distr_C_X_} and~\ref{y0_c0_s1_distr_C_X__grouped} illustrate the 
distributions for the $C_{TC}$ (tail complement) area. It seems that $C_{TC}$ has the same 
distribution as $C_T$ for all samples except for the sample ``Productive", for which
$C_{TC}$ has slightly higher values than~$C_T$ does. Note, also, that the ``Productive" distribution 
has lower values for \hi than ``Top h". This means that the height of the $C_{TC}$ areas is 
smaller for the ``Productive" authors than for ``Top H' ones. The previous two remarks lead to 
the (rather expected) conclusion that the ``Productive" authors have long and thin tails.

The $C_{IC}$ distribution is shown in Figures~\ref{y0_c0_s1_distr_C_I_} 
and~\ref{y0_c0_s1_distr_C_I__grouped}. In these plots, it is clear that the ``Productive" 
authors have clearly higher values than any other sample, since $C_{IC}$ is strongly related 
with the total number of publications.

\begin{figure}[!tb]
\centering
\subfigure[$PI$ (* 10000)]{\label{y0_c0_s1_distr_PI_} \resizebox{5.5cm}{3.8cm}{ \psfig{figure=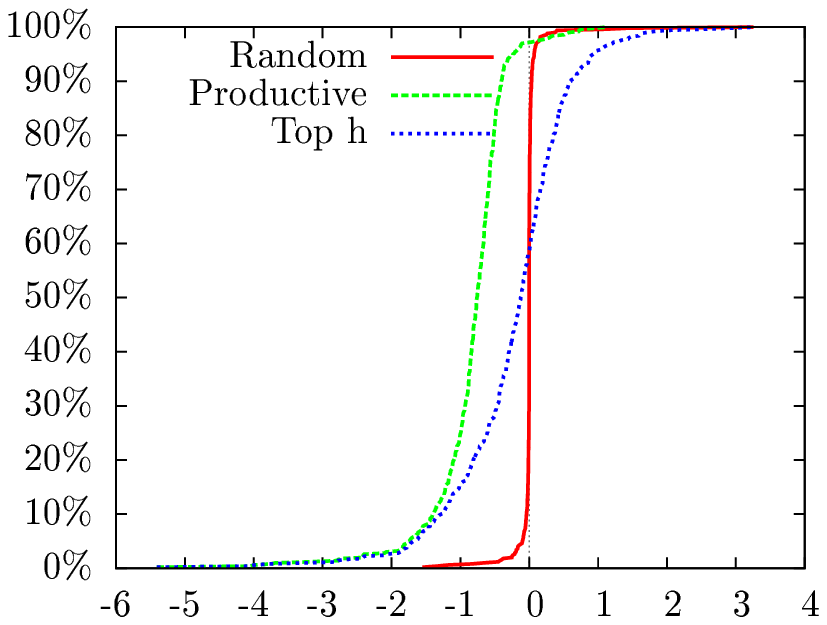,clip=}}}
\subfigure[$PI$ (* 10000)]{\label{y0_c0_s1_distr_PI__grouped} \resizebox{5.5cm}{3.8cm}{ \psfig{figure=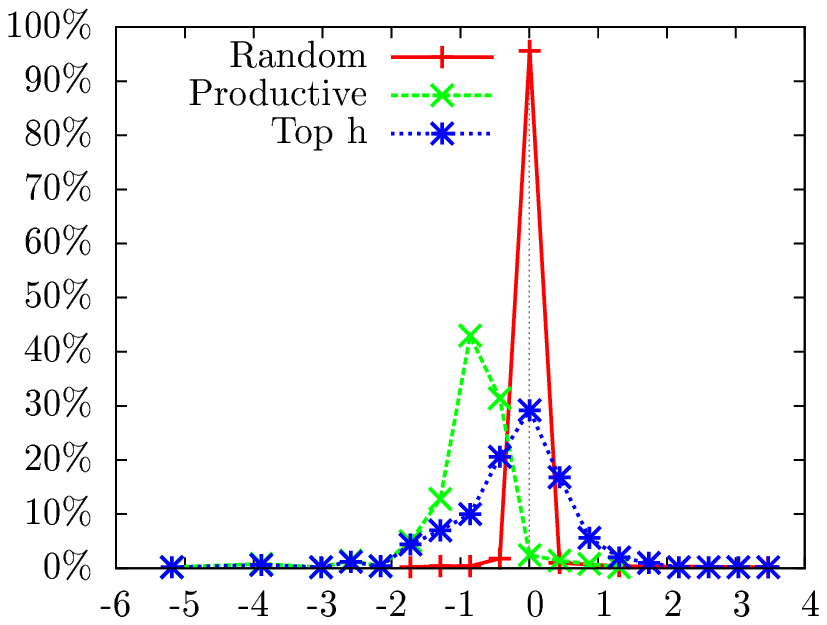,clip=}}}
\subfigure[$PI_{(\kappa=2)}$ (* 10000)]{\label{y0_c0_s1_distr_PI_k2_} \resizebox{5.5cm}{3.8cm}{ \psfig{figure=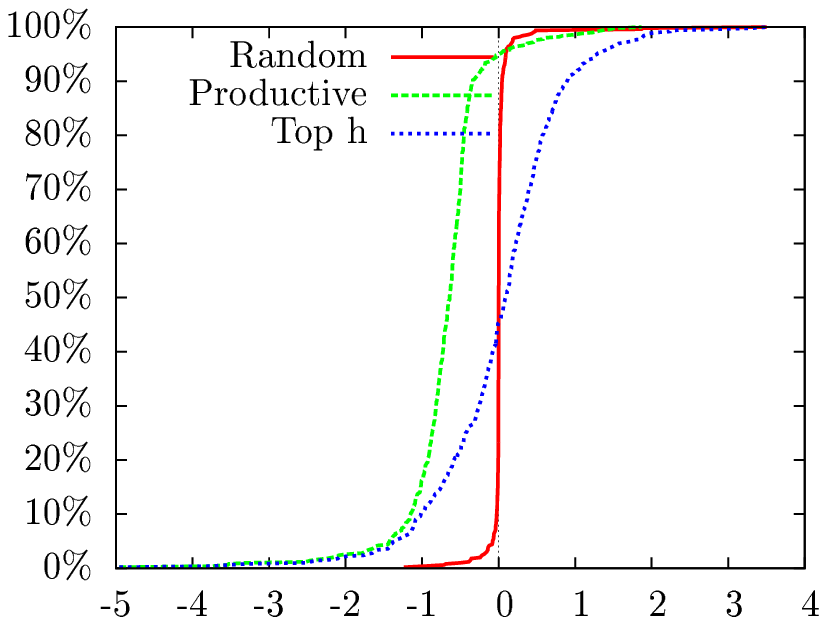,clip=}}}
\subfigure[$PI_{(\kappa=2)}$ (* 10000)]{\label{y0_c0_s1_distr_PI_k2__grouped} \resizebox{5.5cm}{3.8cm}{ \psfig{figure=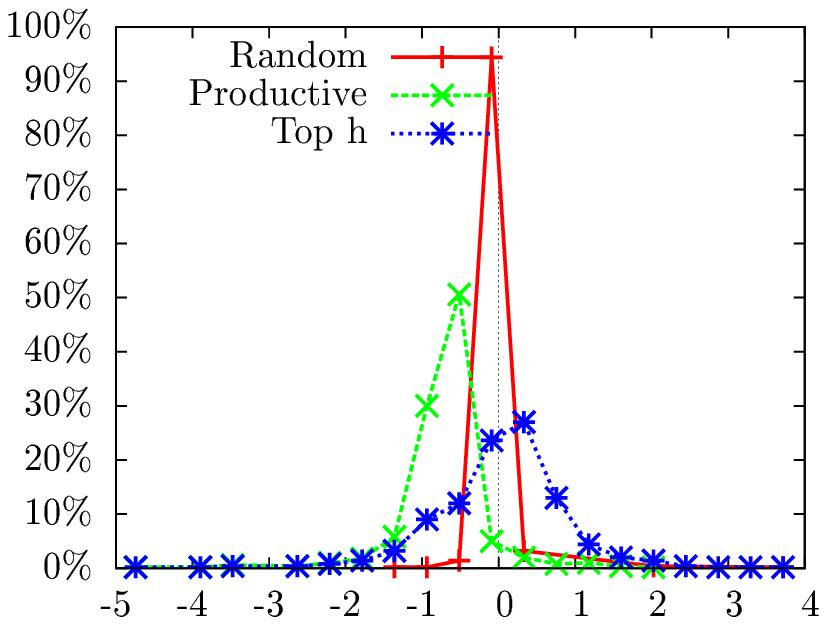,clip=}}}
\subfigure[$PI_{\kappa=4}$ (* 10000)]{\label{y0_c0_s1_distr_PI_k4_} \resizebox{5.5cm}{3.8cm}{ \psfig{figure=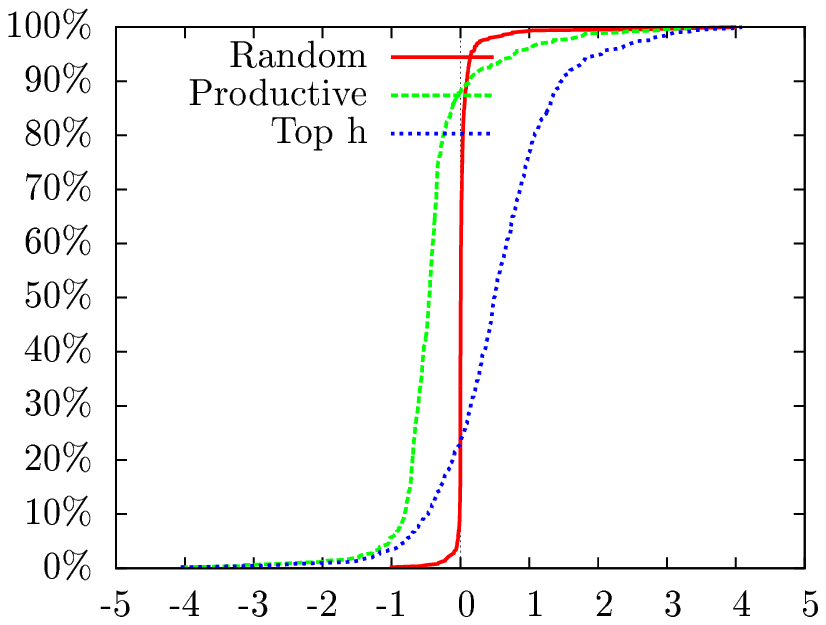,clip=}}}
\subfigure[$PI_{\kappa=4}$ (* 10000)]{\label{y0_c0_s1_distr_PI_k4__grouped} \resizebox{5.5cm}{3.8cm}{ \psfig{figure=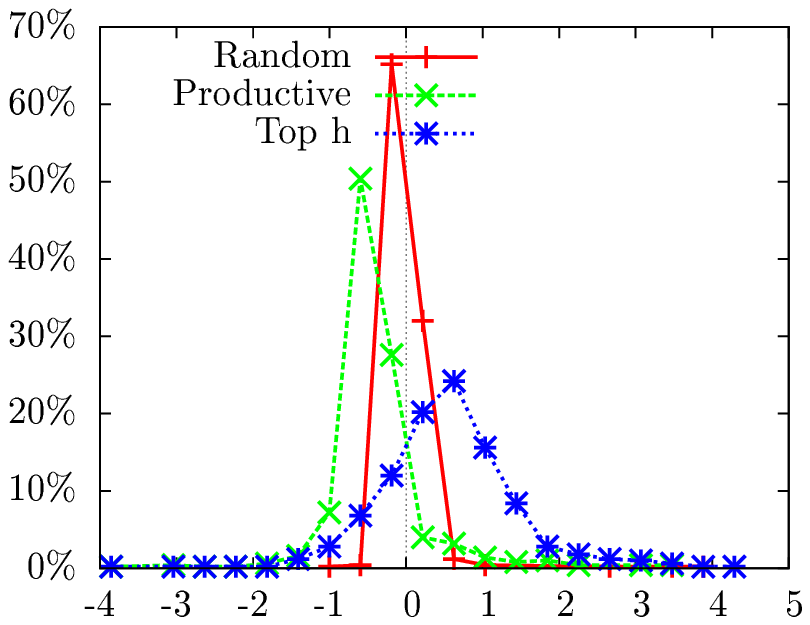,clip=}}}
\subfigure[$PI$ (limited x range)]{\label{y0_c0_s1_distr_PI__-500_500} \resizebox{5.5cm}{3.8cm}{ \psfig{figure=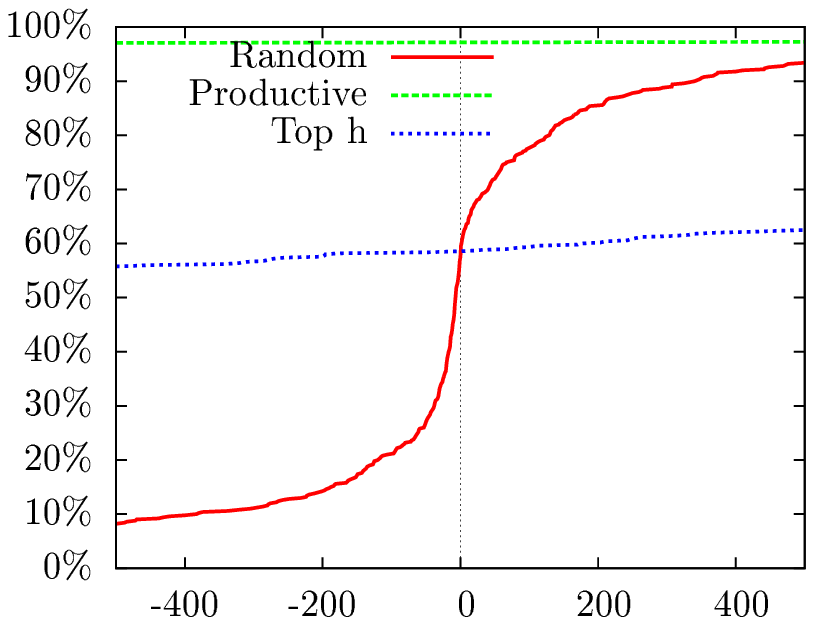,clip=}}}
\subfigure[$PI_{\kappa=4}$ (limited x range)]{\label{y0_c0_s1_distr_PI_k4__-500_500} \resizebox{5.5cm}{3.8cm}{ \psfig{figure=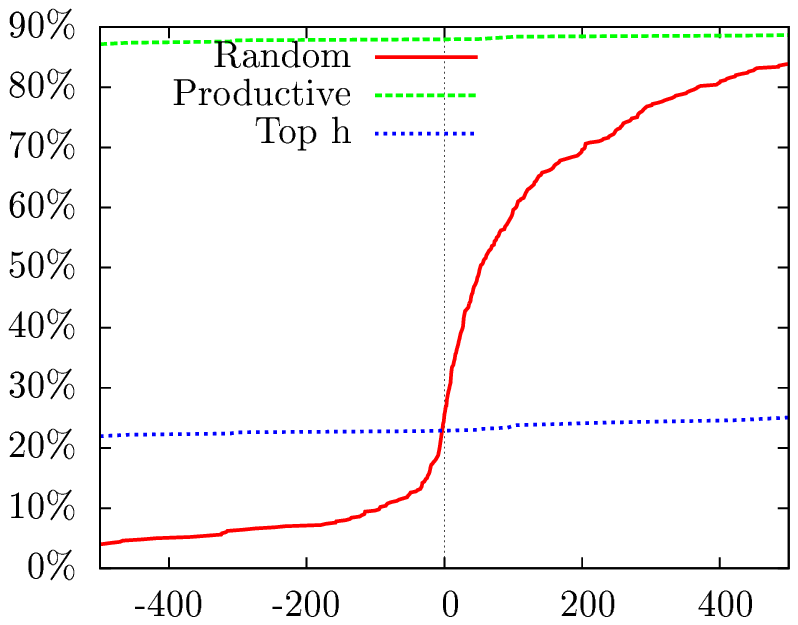,clip=}}}
\caption{Distributions of $PI$, $PI_{(\kappa=2)}$ and $PI_{(\kappa=4)}$}
\label{fig_plot_all_plot_sh_vals1}
\end{figure}

In Figure~\ref{fig_plot_all_plot_sh_vals1} we see the distributions for the previously 
defined $PI$ index. Interestingly, it seems that the value~0 is a key value. For all plots, 
the zero y-axis is the center of the figure. As seen in Figures~\ref{y0_c0_s1_distr_PI_} 
and~\ref{y0_c0_s1_distr_PI__grouped} most of the authors are located around zero. Note that 
in the right plots, a point at $x=0, y=95\%$ with a previous value of $x=-3000$ means that the 
95\% of the authors have values in the range $-1500 \dots 1500$. The first two plots 
show that the ``Top h" authors have the highest values for $PI$ (about 10\% of them have values 
greater than~8000). 

Figure~\ref{y0_c0_s1_distr_PI__-500_500} is a zoomed-in version of Figure~\ref{y0_c0_s1_distr_PI_}. 
It is clear that about 96\% of the ``Productive" authors have~$PI<0$. This means that in this 
sample there are a lot of ``mass producers" (people with high number of publications but 
relatively low \hi - or at least not in ``Top h-indexers"). The other samples cut the zero 
y-axis at about 50\% to 60\%, which means that 40\% to 50\% are positive. It is also noticeable 
that about 70\% (15-85\%) of the ``Random sample have values very close to zero within the range 
-200..200.

In Figure~\ref{y0_c0_s1_distr_PI_k2_}, \subref{y0_c0_s1_distr_PI_k2__grouped}, 
\subref{y0_c0_s1_distr_PI_k4_} and~\subref{y0_c0_s1_distr_PI_k4__grouped} we present the 
distributions for $PI_{\kappa=2}$ and $PI_{\kappa=4}$. We remind that factor $\kappa$ is the 
core area multiplier. In these plots, it is shown that these distributions behave like the 
basic $PI$ distribution except that they are slightly shifted to the right. 
The "Productive" sample is affected less than the others. 
This outcome is understandable since they are the authors with small \hi core areas 
compared to their tail and excess areas.

Comparing subfigure~\subref{y0_c0_s1_distr_PI_k4__-500_500} 
to~\subref{y0_c0_s1_distr_PI__-500_500} we can better see the differences. The number of 
authors in the negative side of samples ``Random" and ``Top h" has decreased from 
57\% and 58\% to 24\% and 23\% respectively, 
meaning that about 33\%-35\% of the sample members moved from the negative to the positive 
side. The number of ``Productive" authors in the negative side has been decreased from 97\% 
to 88\%, i.e. an additional 11\% of the sample members moved to the positive side.

In addition to the distribution plots, Table~\ref{positive_tab} presents the number of authors 
that have the mentioned metrics below or above zero for each sample. As mentioned before, 97\% 
of the ``Productive" authors have $PI<0$, whereas only 3\% reside in the positive side of 
the plot. This amount increases as we increase the core factor $\kappa$. For $\kappa=4$ the 
increment is 9\% (12\% from 3\%). In all other samples the increment is greater, i.e. for 
``Top h" the increment is 35\%, for ``Random" is 33\%.

\begin{figure}[!tb]
\centering
\subfigure[$XPI$ (* 100000)]{\label{y0_c0_s1_distr_XPI_} \resizebox{5.5cm}{3.8cm}{ \psfig{figure=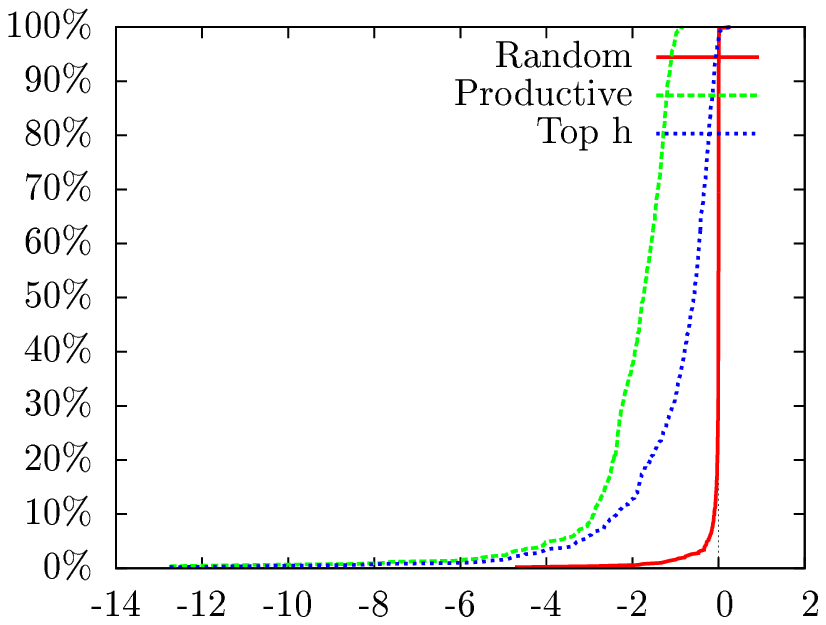,clip=}}}
\subfigure[$XPI$ (* 100000)]{\label{y0_c0_s1_distr_XPI__grouped} \resizebox{5.5cm}{3.8cm}{ \psfig{figure=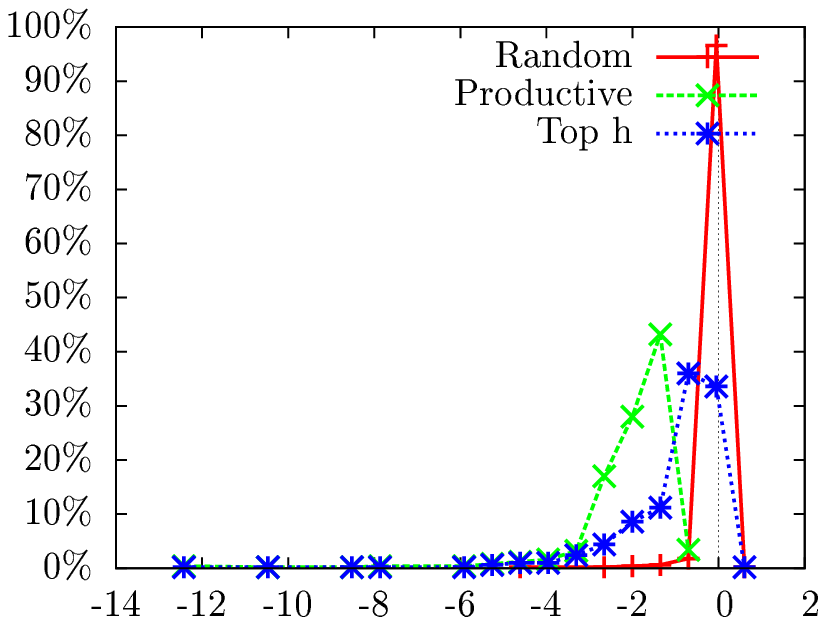,clip=}}}
\subfigure[$XPI$ (limited x range)]{\label{y0_c0_s1_distr_XPI__-500_500} \resizebox{5.5cm}{3.8cm}{ \psfig{figure=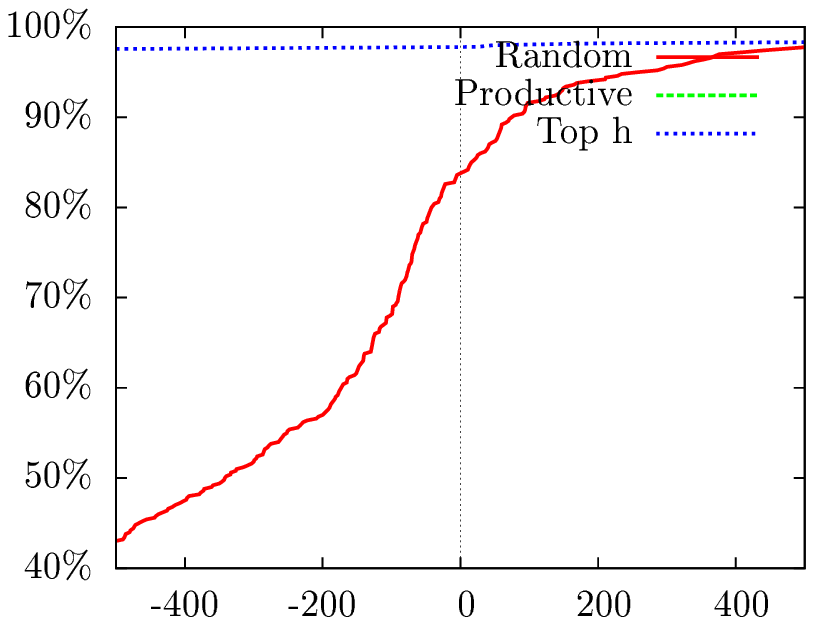,clip=}}}
\caption{Distributions of $XPI$}
\label{fig_plot_all_plot_vsh_vals1}
\end{figure}

In Figure~\ref{fig_plot_all_plot_vsh_vals1} the same kinds of plots are presented for the metric 
$XPI$. As expected, the difference is that most of the authors lie in the negative side of the 
graph. The cut points of y-axis are also presented in Table~\ref{positive_tab}. About 2\% of 
the ``Top \hi" authors have $XPI>0$ but {\bf none} of the ``Productive" authors do. The cut point 
for ``Random" authors is at 6\%. Also, at this point we repeat the experiment of varying the 
$\kappa$ value. The results do not match with those of the $PI$ case. Incrementing $\kappa$ does 
not increase the number of positive authors in the same way as the $PI$ case. The increment 
is negligible for the ``Productive" and ``Top h" and very small for the sample ``Random". This 
leads to the conclusion that varying the $\kappa$ factor does not affect $XPI$ significantly. 
Probably different default values for the factors of Equation~\ref{eq_PI} (especially for 
$\kappa$ and/or $\nu$) may be needed for tuning the $XPI$ metric. However, this task remains 
out of the scope of the present article.

\begin{table}[!tb]
\caption{$PI$ and $XPI$ statistics.}
\label{positive_tab}
\centering
\footnotesize 
\begin{tabular}{|r@{\hspace{0.3em}}|@{\hspace{0.3em}}cc@{\hspace{0.3em}}|@{\hspace{0.3em}}cc@{\hspace{0.3em}}|@{\hspace{0.3em}}cc@{\hspace{0.3em}}|@{\hspace{0.3em}}cc@{\hspace{0.3em}}|@{\hspace{0.3em}}cc@{\hspace{0.3em}}|@{\hspace{0.3em}}cc@{\hspace{0.3em}}|}\hline
\multirow{2}{*}{\vspace{0.1cm} Sample} &\multicolumn{2}{@{\hspace{-0.3em}}|c@{\hspace{-0.3em}}}{$PI$} &\multicolumn{2}{@{\hspace{-0.3em}}|c}{$PI_{\kappa=2}$} &\multicolumn{2}{@{\hspace{-0.3em}}|c}{$PI_{\kappa=4}$} &\multicolumn{2}{@{\hspace{-0.3em}}|c}{$XPI$} &\multicolumn{2}{@{\hspace{-0.3em}}|c}{$XPI_{\kappa=2}$} &\multicolumn{2}{@{\hspace{-0.3em}}|c|}{$XPI_{\kappa=4}$}\\
 & $<0$ & $\ge 0$ & $<0$ & $\ge 0$ & $<0$ & $\ge 0$ & $<0$ & $\ge 0$ & $<0$ & $\ge 0$ & $<0$ & $\ge 0$\\ \hline\hline
\multirow{2}{*}{Random} & 284 & 216 & 213 & 287 & 122 & 378 & 418 & 82 & 408 & 92 & 383 & 117\\
 & 57\% &43\% & 43\% &57\% & 24\% &76\% & 84\% &16\% & 82\% &18\% & 77\% &23\%\\\hline
\multirow{2}{*}{Productive} & 485 & 15 & 474 & 26 & 439 & 61 & 500 & 0 & 500 & 0 & 500 & 0\\
 & 97\% &3\% & 95\% &5\% & 88\% &12\% & 100\% &0\% & 100\% &0\% & 100\% &0\%\\\hline
\multirow{2}{*}{Top $h$} & 292 & 208 & 226 & 274 & 114 & 386 & 488 & 12 & 484 & 16 & 473 & 27\\
 & 58\% &42\% & 45\% &55\% & 23\% &77\% & 98\% &2\% & 97\% &3\% & 95\% &5\%\\\hline
\multirow{2}{*}{Unioned} & 904 & 419 & 767 & 556 & 563 & 760 & 1230 & 93 & 1216 & 107 & 1180 & 143\\
 & 68\% &32\% & 58\% &42\% & 43\% &57\% & 93\% &7\% & 92\% &8\% & 89\% &11\%\\\hline
\end{tabular}
\normalsize

\end{table}

\subsection{$PI$ robustness to self-citations}
\label{subsec:self}

Self-citations (a citation from an article to another article when there is at least one 
common author between the citing and the cited paper) is a common way for authors to increase 
the visibility of their works; it has significant impact on \hi~\cite{Schreiber-EPL07}, and 
some efforts has been made towards designing robust metrics to 
self-citations~\cite{Katsaros-JASIST09}.

\begin{figure}[!b]
\centering
\subfigure[$h$ vs. $h$(no-self)]{\label{qqplots_hindex_c0_s1_y0_vs_hindex_c0_s0_y0_all} \resizebox{5.5cm}{3.8cm}{ \psfig{figure=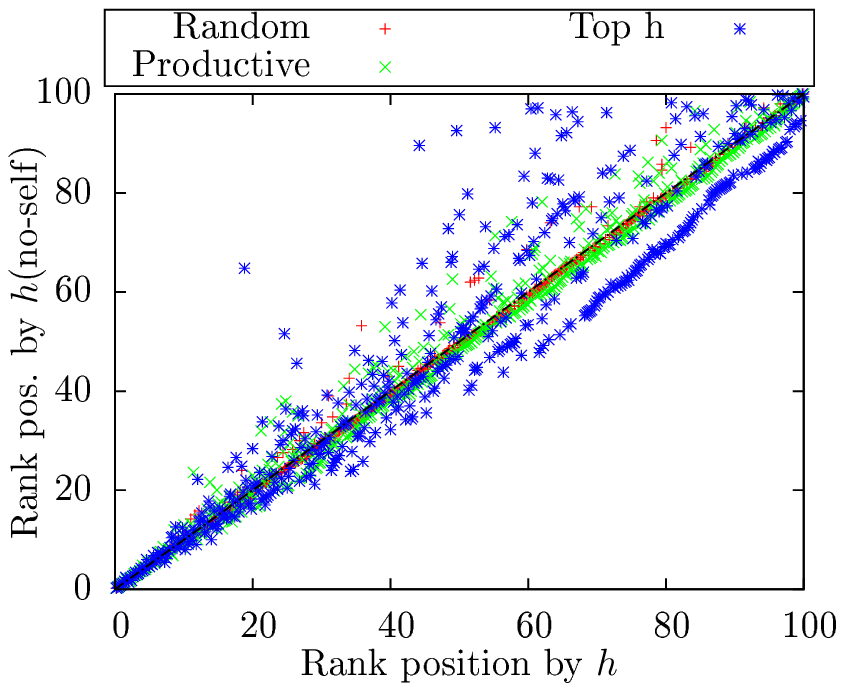,clip=}}}
\subfigure[$PI$ vs. $PI$(no-self)]{\label{qqplots_PI_c0_s1_y0_vs_PI_c0_s0_y0_all} \resizebox{5.5cm}{3.8cm}{ \psfig{figure=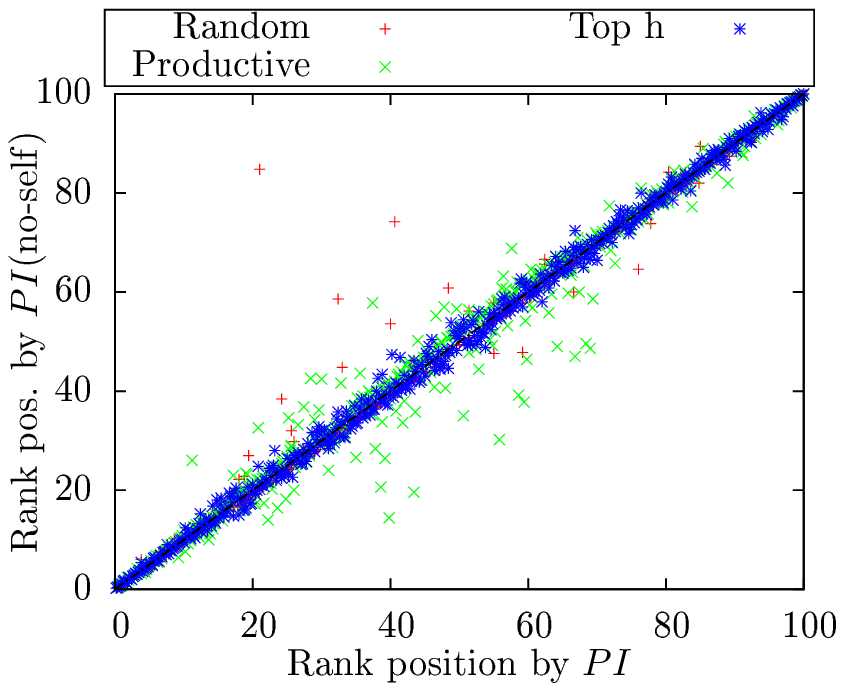,clip=}}}
\caption{Robustness of \hi and PI to self-citations (Q-Q plots: X and Y axis denote the rank position normalized in percent).}
\label{plot_all_qqplot_self}
\end{figure}

We performed an experiment to study the behavior of \hi and PI with respect to self-citations. 
In Figure~\ref{qqplots_hindex_c0_s1_y0_vs_hindex_c0_s0_y0_all} a qq-plot is shown, which 
compares the ranking produced by \hi. The x-axis represents the rank produced by the 
computed \hi including self-citations, whereas the y-axis represents the rank of \hi after 
excluding self-citations. We have performed several experiments with different types 
of ranking and they all show similar behavior with respect to \hi. In Figure~\ref{qqplots_PI_c0_s1_y0_vs_PI_c0_s0_y0_all} the same kind of qq-plot for the $PI$ as 
a rank criterion is displayed. It is apparent that $PI$ is much less affected by self-citations 
than the \hi. This is another advantage of the proposed metric; it is not affected by 
self-citations.

\section{PI in action: Ranking scientists}
\label{sec:results}

In the previous two subsections, we performed an analysis of the datasets at a coarse level. 
In this section, we will provide an analysis at a finer level, that of individual 
scientists. We have emphasized from the beginning of the article that it is not this article's 
purpose to explain the roots of the publishing behaviour of individual scientists. However, 
we will attempt to record those characteristics of the scientists (if there are such 
characteristics) that make them exhibit particular behaviours.

\renewcommand{\arraystretch}{0.92}%

\begin{table}[!bt]
\caption{Ranking by \hi (top-20 scientists).}
\label{tab_rank_table_hindex}
\begin{center}
\footnotesize
\begin{tabular}{|l@{}||@{\hspace{.1cm}}rr@{\hspace{.1cm}}|@{\hspace{.1cm}}rr@{\hspace{.1cm}}|r|r|r|r|}\hline
\multirow{2}{*}{\bf Author} 	&\multicolumn{2}{@{\hspace{-.1cm}}|c@{\hspace{.1cm}}}{\bf $h$} &\multicolumn{2}{@{\hspace{-.1cm}}|c@{\hspace{.1cm}}}{\bf $PI$} & \multicolumn{1}{|c|}{\multirow{2}{*}{\bf $p$}}& \multicolumn{1}{|c|}{\multirow{2}{*}{\bf $C$}}& \multicolumn{1}{|c|}{\multirow{2}{*}{\bf $C/p$}} &\multicolumn{1}{|c@{\hspace{.1cm}}|}{change} \\
		& val&pos	&val	& pos &     &       &      & $h$-$PI$\\\hline\hline
Shenker Scott	& 97 & 1	& 5754 	& 52  & 508 & 45621 & 89.81 & -51 	\\ 
Foster Ian	& 93 & 2	& -15510 	& 1287& 768 & 47265 & 61.54 & -1285 	\\ 
Garcia-Molina Hector& 92 & 3	& -17423 	& 1299& 605 & 29773 & 49.21 & -1296 	\\ 
Estrin Deborah	& 90 & 4	& 5348 	& 62  & 479 & 40358 & 84.25 & -58 	\\ 
Ullman Jeffrey	& 86 & 5	& 11267 	& 18  & 460 & 43431 & 94.42 & -13 	\\ 
Culler David	& 84 & 6	& 7552 	& 38  & 386 & 32920 & 85.28 & -32 	\\ 
Tarjan Robert	& 83 & 7	& 2888 	& 117 & 405 & 29614 & 73.12 & -110 	\\ 
Towsley Don	& 82 & 8	& -31929 	& 1318& 793 & 26373 & 33.26 & -1310 	\\ 
Kanade T.	& 81 & 9	& -20753 	& 1309& 742 & 32788 & 44.19 & -1300 	\\ 
Haussler David	& 81 & 10	& 10952 	& 19  & 335 & 31526 & 94.11 & -9 	\\ 
Jain Anil	& 81 & 11	& -11474 	& 1236& 590 & 29755 & 50.43 & -1225 	\\ 
Papadimitriou Christos&80& 12	& -5897 	& 968 & 506 & 28183 & 55.70 & -956 	\\ 
Katz Randy	& 78 & 13	& -27820 	& 1317& 757 & 25142 & 33.21 & -1304 	\\ 
Pentland Alex	& 77 & 14	& -1242 	& 724 & 509 & 32022 & 62.91 & -710 	\\ 
Han Jiawei	& 77 & 15	& -15410 	& 1285& 653 & 28942 & 44.32 & -1270 	\\ 
Jordan Michael	& 75 & 16	& -1062 	& 717 & 499 & 30738 & 61.60 & -701 	\\ 
Karp Richard	& 75 & 17	& 7231 	& 41  & 377 & 29881 & 79.26 & -24 	\\ 
Zisserman A.	& 75 & 18	& 210 	& 263 & 421 & 26160 & 62.14 & -245 	\\ 
Jennings Nick	& 74 & 19	& -15718 	& 1289& 626 & 25130 & 40.14 & -1270 	\\ 
Thrun S.	& 74 & 20	& -5789 	& 958 & 445 & 21665 & 48.69 & -938 	\\ \hline
\end{tabular}
\normalsize

\end{center}
\end{table}

Table~\ref{tab_rank_table_hindex} shows the rank table for the top-20 authors by \hi from all our 
samples. They are truly remarkable scientists with significant contributions to their field. 
The table also shows their corresponding $PI$ values; it is remarkable that about half of them 
are characterized as ``Mass Producers" (i.e., they have negative $PI$ values). We will seek an 
explanation for that by contrasting these results in  Table~\ref{tab_rank_table_PI}.

\begin{table}[!t]
\caption{Ranking by $PI$ index (top-20 influential scientists).}
\label{tab_rank_table_PI}
\begin{center}
\footnotesize
\begin{tabular}{|l@{}||@{\hspace{.1cm}}rr@{\hspace{.1cm}}|@{\hspace{.1cm}}rr@{\hspace{.1cm}}|r|r|r|r|}\hline
\multirow{2}{*}{\bf Author } &\multicolumn{2}{@{\hspace{-.1cm}}|c@{\hspace{.1cm}}}{\bf $PI$} &\multicolumn{2}{@{\hspace{-.1cm}}|c@{\hspace{.1cm}}}{\bf $h$} & \multicolumn{1}{|c|}{\multirow{2}{*}{\bf $p$}}& \multicolumn{1}{|c|}{\multirow{2}{*}{\bf $C$}}& \multicolumn{1}{|c|}{\multirow{2}{*}{\bf $C/p$}} &\multicolumn{1}{|c@{\hspace{.1cm}}|}{change} \\
		& val   & pos& val& pos &     &       &        & $h$-$PI$\\\hline\hline
Vapnik Vladimir& 32542 & 1  & 50 & 171 & 126 & 36342 & 288.43 & +170 	\\ 
Rivest Ronald	& 29340 & 2  & 62 & 53  & 320 & 45336 & 141.68 & +51 	\\ 
Zadeh L.	& 25613 & 3  & 59 & 70  & 320 & 41012 & 128.16 & +67 	\\ 
Kohonen Teuvo	& 19880 & 4  & 51 & 157 & 160 & 25439 & 158.99 & +153 	\\ 
Floyd Sally	& 18059 & 5  & 66 & 38  & 222 & 28355 & 127.73 & +33 	\\ 
Kesselman Carl	& 17054 & 6  & 60 & 64  & 272 & 29774 & 109.46 & +58 	\\ 
Schapire Robert& 16169 & 7  & 56 & 90  & 186 & 23449 & 126.07 & +83 	\\ 
Milner Robin	& 16019 & 8  & 54 & 108 & 202 & 24011 & 118.87 & +100 	\\ 
Shamir A	& 15926 & 9  & 53 & 125 & 213 & 24406 & 114.58 & +116 	\\ 
Tuecke Steven	& 14747 & 10 & 44 & 281 & 96  & 17035 & 177.45 & +271 	\\ 
Balakrishnan Hari& 14444 & 11 & 72 & 21  & 272 & 28844 & 106.04 & +10 	\\ 
Agrawal Rakesh	& 14375 & 12 & 67 & 30  & 353 & 33537 & 95.01  & +18 	\\ 
Hinton G.	& 13415 & 13 & 63 & 45  & 314 & 29228 & 93.08  & +32 	\\ 
Aho Alfred	& 13048 & 14 & 50 & 173 & 193 & 20198 & 104.65 & +159 	\\ 
Lamport Leslie	& 12254 & 15 & 59 & 71  & 273 & 24880 & 91.14  & +56 	\\ 
Hopcroft John	& 12088 & 16 & 45 & 258 & 198 & 18973 & 95.82  & +242 	\\ 
Morris Robert	& 11685 & 17 & 57 & 81  & 305 & 25821 & 84.66  & +64 	\\ 
Ullman Jeffrey	& 11267 & 18 & 86 & 5   & 460 & 43431 & 94.42  & -13 	\\ 
Haussler David	& 10952 & 19 & 81 & 10  & 335 & 31526 & 94.11  & -9 	\\ 
Joachims T.	& 10767 & 20 & 41 & 377 & 134 & 14580 & 108.81 & +357 	\\ \hline
\end{tabular}
\normalsize
\end{center}
\end{table}

Table~\ref{tab_rank_table_PI} shows the rank list ordered by $PI$; all authors have high ranking 
positions by \hi as well. If we try to find what is common in all these persons, we could 
say that (most of) these scientists spend significant time of their careers in industrial 
environments making groundbreaking contributions, and being recognized as inventors whose ideas 
have been incorporated into many products that penetrated our lives. Examples include Tuecke, 
Rivest, Shamir, Agrawal, and Lamport. The personnel in these environments are highly trained, 
working on ``real" problems whose solutions are part of business products. Thus, these groups 
are not publishing-prone, and (most of the time) whenever they publish their results, these 
are path-breaking and influential. Others, such as Vapnik, Zadeh, Kohonen, Aho and Schapire 
are pioneers, inventing brand new knowledge and developing it in a long series of articles. 
It might also be the case that these scientists work only with experienced researchers, thus 
being {\it elitists}~\cite{Cormode-JNL-Informetrics13}, because for instance their topics are very 
advanced. On the contrary, people coming solely from academic environments have the role of 
a mentor~\cite{Cormode-JNL-Informetrics13} and are charged with the task of training young 
PhD students whose initial works (usually) do not have high impact. Moreover, sometimes they 
are involved in projects of exploratory nature, which eventually do not open new avenues. 
Finally, we should not forget the publish-or-perish pressure upon their students and themselves.

\begin{table}[!t]
\caption{Ranking by $PI$ (bottom-20 by PI, i.e., top-20 mass producers).}
\label{tab_rank_table_PI_last20}
\begin{center}
\footnotesize
\begin{tabular}{|l@{}||@{\hspace{.1cm}}rr@{\hspace{.1cm}}|@{\hspace{.1cm}}rr@{\hspace{.1cm}}|r|r|r|r|}\hline
\multirow{2}{*}{\bf Author } &\multicolumn{2}{@{\hspace{-.1cm}}|c@{\hspace{.1cm}}}{\bf $PI$} &\multicolumn{2}{@{\hspace{-.1cm}}|c@{\hspace{.1cm}}}{\bf $h$} & \multicolumn{1}{|c|}{\multirow{2}{*}{\bf $p$}}& \multicolumn{1}{|c|}{\multirow{2}{*}{\bf $C$}}& \multicolumn{1}{|c|}{\multirow{2}{*}{\bf $C/p$}} &\multicolumn{1}{|c@{\hspace{.1cm}}|}{change} \\
		& val    & pos  & val& pos &     &       &       & $h$-$PI$\\\hline\hline
Ikeuchi Katsushi	& -18173 & 1303 & 43 & 327 & 638 & 7412  & 11.62 & -976 	\\ 
Thalmann D.	& -18356 & 1304 & 46 & 249 & 632 & 8600  & 13.61 & -1055 	\\ 
Reddy Sudhakar	& -18369 & 1305 & 43 & 322 & 659 & 8119  & 12.32 & -983 	\\ 
Gao Wen		& -18494 & 1306 & 26 & 649 & 907 & 4412  & 4.86  & -657 	\\ 
Prade Henri	& -18692 & 1307 & 65 & 42  & 633 & 18228 & 28.80 & -1265 	\\ 
Liu K.		& -19063 & 1308 & 42 & 355 & 672 & 7397  & 11.01 & -953 	\\ 
Kanade T.	& -20753 & 1309 & 81 & 9   & 742 & 32788 & 44.19 & -1300 	\\ 
Rosenfeld Azriel& -21023 & 1310 & 59 & 73  & 707 & 17209 & 24.34 & -1237 	\\ 
Gupta Anoop	& -23959 & 1311 & 64 & 43  & 739 & 19241 & 26.04 & -1268 	\\ 
Miller J.	& -24112 & 1312 & 40 & 433 & 807 & 6568  & 8.14  & -879 	\\ 
Shin Kang	& -24125 & 1313 & 57 & 85  & 731 & 14293 & 19.55 & -1228 	\\ 
Schmidt Douglas& -24153 & 1314 & 56 & 94  & 729 & 13535 & 18.57 & -1220 	\\ 
Bertino Elisa	& -27058 & 1315 & 49 & 194 & 805 & 9986  & 12.40 & -1121 	\\ 
Yu Philip	& -27727 & 1316 & 63 & 48  & 789 & 18011 & 22.83 & -1268 	\\ 
Katz Randy	& -27820 & 1317 & 78 & 13  & 757 & 25142 & 33.21 & -1304 	\\ 
Towsley Don	& -31929 & 1318 & 82 & 8   & 793 & 26373 & 33.26 & -1310 	\\ 
Kuo C.		& -36848 & 1319 & 40 & 425 & 1148& 7472  & 6.51  & -894 	\\ 
Gerla Mario	& -37464 & 1320 & 67 & 32  & 945 & 21362 & 22.61 & -1288 	\\ 
Dongarra Jack	& -39901 & 1321 & 67 & 31  & 982 & 21404 & 21.80 & -1290 	\\ 
Poor H.		& -40492 & 1322 & 55 & 100 & 1069& 15278 & 14.29 & -1222 	\\ 
Huang Thomas	& -54047 & 1323 & 67 & 33  & 1172& 19988 & 17.05 & -1290 	\\ \hline
\end{tabular}
\normalsize
\end{center}
\end{table}

Table~\ref{tab_rank_table_PI_last20} shows the top-20 ``Mass Producers" from our samples. 
In this table we also present the average number of citations per paper ($C/p$ column). It can 
be seen that there is a big range of average values from 4 to 45 citations per publication in 
the top ``Mass Producers". In this table we will recognize -- consistent with what we said in 
the previous paragraph -- some excellent academics who have trained many PhD students.

\begin{table}[!t]
\caption{Rank table by $PI$  of sample ``Networks".}
\label{tab_rank_table_PI_s45}
\begin{center}
\footnotesize
\begin{tabular}{|l@{}||@{\hspace{.1cm}}rr@{\hspace{.1cm}}|@{\hspace{.1cm}}rr@{\hspace{.1cm}}|r|r|r|r|}\hline
\multirow{2}{*}{\bf Author } &\multicolumn{2}{@{\hspace{-.1cm}}|c@{\hspace{.1cm}}}{\bf $PI$} &\multicolumn{2}{@{\hspace{-.1cm}}|c@{\hspace{.1cm}}}{\bf $h$} & \multicolumn{1}{|c|}{\multirow{2}{*}{\bf $p$}}& \multicolumn{1}{|c|}{\multirow{2}{*}{\bf $C$}}& \multicolumn{1}{|c|}{\multirow{2}{*}{\bf $C/p$}} &\multicolumn{1}{|c@{\hspace{.1cm}}|}{change} \\
		&val    & pos& val& pos &  &  & & $h$-$PI$\\\hline\hline
Jacobson Van	& 19982 & 1  & 44 & 44  & 161 & 25130 & 156.09 & +43 	\\ 
Floyd Sally	& 18059 & 2  & 66 & 9  &  222 & 28355  & 127.73 & +7 	\\ 
Balakrishnan Hari& 14444 & 3  & 72 & 7   & 272 & 28844 & 106.04 & +4 	\\ 
Johnson David	& 12180 & 4  & 54 & 21  & 263 & 23466 & 89.22 & +17 	\\ 
Morris Robert	& 11685 & 5  & 57 & 16  & 305 & 25821 & 84.66 & +11 	\\ 
Handley M.	& 10763 & 6  & 47 & 35  & 201 & 18001 & 89.56 & +29 	\\ 
Perkins C.	& 9609  & 7  & 52 & 25  & 373 & 26301 & 70.51 & +18 	\\ 
Paxson Vern	& 8871  & 8  & 60 & 12  & 233 & 19251 & 82.62 & +4 	\\ 
Stoica Ion	& 8558  & 9  & 63 & 11  & 266 & 21347 & 80.25 & +2 	\\ 
Heidemann John	& 8059  & 10 & 47 & 36  & 237 & 16989 & 71.68 & +26 	\\ 
Culler David	& 7552  & 11 & 84 & 3   & 386 & 32920 & 85.28 & -8 	\\ 
Shenker Scott	& 5754  & 12 & 97 & 1   & 508 & 45621 & 89.81 & -11 	\\ 
Govindan Ramesh& 5356  & 13 & 55 & 19  & 287 & 18116 & 63.12 & +6 	\\ 
Estrin Deborah	& 5348  & 14 & 90 & 2   & 479 & 40358 & 84.25 & -12 	\\ 
Crovella Mark	& 4886  & 15 & 46 & 38  & 172 & 10682 & 62.10 & +23 	\\ 
Perrig Adrian	& 4304  & 16 & 58 & 14  & 247 & 15266 & 61.81 & -2 	\\ 
Lu Songwu	& 3430  & 17 & 44 & 45  & 129 & 7170  & 55.58 & +28 	\\ 
Akyildiz Ian	& 3089  & 18 & 53 & 23  & 401 & 21533 & 53.70 & +5 	\\ 
Kleinrock Leonard& 1986  & 19 & 51 & 31  & 282 & 13767 & 48.82 & +12 	\\ 
Knightly Edward& 263   & 20 & 41 & 50  & 172 & 5634  & 32.76 & +30 	\\ 
Peterson L.	& -652  & 21 & 54 & 22  & 292 & 12200 & 41.78 & +1 	\\ 
Hubaux Jean-Pierre& -653  & 22 & 45 & 43  & 247 & 8437  & 34.16 & +21 	\\ 
Vaidya Nitin	& -1242 & 23 & 50 & 32  & 337 & 13108 & 38.90 & +9 	\\ 
Zhang Lixia	& -1609 & 24 & 55 & 20  & 374 & 15936 & 42.61 & -4 	\\ 
Low Steven	& -1796 & 25 & 45 & 42  & 291 & 9274  & 31.87 & +17 	\\ 
Boudec Jean-Yves& -2338 & 26 & 44 & 46  & 258 & 7078  & 27.43 & +20 	\\ 
Win Moe		& -2619 & 27 & 46 & 37  & 341 & 10951 & 32.11 & +10 	\\ 
Rexford Jennifer& -2632 & 28 & 49 & 33  & 269 & 8148  & 30.29 & +5 	\\ 
Zhang Hui	& -3344 & 29 & 52 & 27  & 352 & 12256 & 34.82 & -2 	\\ 
Srikant R.	& -3827 & 30 & 46 & 39  & 328 & 9145  & 27.88 & +9 	\\ 
Diot Christophe& -4054 & 31 & 52 & 30  & 290 & 8322  & 28.70 & -1 	\\ 
Simon Marvin	& -4450 & 32 & 42 & 49  & 370 & 9326  & 25.21 & +17 	\\ 
Ammar Mostafa	& -4547 & 33 & 43 & 48  & 308 & 6848  & 22.23 & +15 	\\ 
Kurose Jim	& -5114 & 34 & 59 & 13  & 391 & 14474 & 37.02 & -21 	\\ 
Campbell Andrew& -6036 & 35 & 46 & 41  & 348 & 7856  & 22.57 & +6 	\\ 
Chlamtac I.	& -6274 & 36 & 43 & 47  & 357 & 7228  & 20.25 & +11 	\\ 
Crowcroft Jon	& -6863 & 37 & 48 & 34  & 404 & 10225 & 25.31 & -3 	\\ 
Whitt W.	& -7759 & 38 & 52 & 29  & 394 & 10025 & 25.44 & -9 	\\ 
Goldsmith A.	& -7819 & 39 & 57 & 17  & 479 & 16235 & 33.89 & -22 	\\ 
Srivastava Mani& -8139 & 40 & 57 & 18  & 423 & 12723 & 30.08 & -22 	\\ 
Paulraj A.	& -8421 & 41 & 64 & 10  & 442 & 15771 & 35.68 & -31 	\\ 
Schulzrinne Henning& -11050& 42 & 53 & 24  & 555 & 15556 & 28.03 & -18 	\\ 
Garcia-Luna-Aceves J.& -11169& 43& 46 & 40  & 460 & 7875  & 17.12 & -3 	\\ 
Nahrstedt Klara& -12286& 44 & 52 & 28  & 492 & 10594 & 21.53 & -16 	\\ 
Cioffi J.	& -14685& 45 & 52 & 26  & 575 & 12511 & 21.76 & -19 	\\ 
Mukherjee B.	& -15702& 46 & 58 & 15  & 535 & 11964 & 22.36 & -31 	\\ 
Katz Randy	& -27820& 47 & 78 & 5   & 757 & 25142 & 33.21 & -42 	\\ 
Towsley Don	& -31929& 48 & 82 & 4   & 793 & 26373 & 33.26 & -44 	\\ 
Gerla Mario	& -37464& 49 & 67 & 8   & 945 & 21362 & 22.61 & -41 	\\ 
Giannakis Georgios& -44707& 50 & 77 & 6   & 932 & 21128 & 22.67 & -44 	\\ \hline
\end{tabular}
\normalsize

\end{center}
\end{table}

In Tables~\ref{tab_rank_table_PI_s45} to~\ref{tab_rank_table_PI_s35} we present the ``toppers" 
with respect to the fields of ``Networks", ``Databases" and ``Multimedia", respectively. 
Starting from the ``Networks' table, we see Van Jacobson and Sally Floyd ranked first and second 
respectively; they are well-known inventors who contributed fundamental algorithms to the design 
and operation of the Internet. They both had careers in industry: in Cisco, Xerox, AT\&T Center 
for Internet Research at ICSI, and worked extensively on developing standards (RFC) in the areas 
of TCP/IP congestion control. Similarly, looking at Table~\ref{tab_rank_table_PI_s25} for the 
database field, we will find in the top positions persons such as Rakesh Agrawal, Ronald Fagin, 
Jeffrey Ullman and Rajeen Motwani who also have spent their careers in companies such as Google, 
IBM and Microsoft, or have contributed fundamental algorithms in fields such as compilers, 
databases and algorithms. We can make similar observations from Table~\ref{tab_rank_table_PI_s35} 
where we find some entrepreneurs such as Ramesh Jain who founded or co-founded multiple startup 
companies including Imageware, Virage and Praja. The type of career is certainly a factor that 
helps categorize a scientist as an influential, since we can see that Van Jacobson (``Networks") 
and Nick Koudas (from ``Databases" who spend part of his career in AT\&T) are the ones who 
gained the greatest rise in PI ranking compared to the h-ranking: 43 and 38 positions, 
respectively. 

\begin{table}[!t]
\caption{Rank table by $PI$  of sample ``DataBases".}
\label{tab_rank_table_PI_s25}
\begin{center}
\footnotesize
\begin{tabular}{|l@{}||@{\hspace{.1cm}}rr@{\hspace{.1cm}}|@{\hspace{.1cm}}rr@{\hspace{.1cm}}|r|r|r|r|}\hline
\multirow{2}{*}{\bf Author } &\multicolumn{2}{@{\hspace{-.1cm}}|c@{\hspace{.1cm}}}{\bf $PI$} &\multicolumn{2}{@{\hspace{-.1cm}}|c@{\hspace{.1cm}}}{\bf $h$} & \multicolumn{1}{|c|}{\multirow{2}{*}{\bf $p$}}& \multicolumn{1}{|c|}{\multirow{2}{*}{\bf $C$}}& \multicolumn{1}{|c|}{\multirow{2}{*}{\bf $C/p$}}&\multicolumn{1}{|c@{\hspace{.1cm}}|}{change} \\
		& val   & pos& val& pos&     &       &       & $h$-$PI$ \\ \hline\hline
Agrawal Rakesh	& 14375 & 1  & 67 & 8  & 353 & 33537 & 95.01 & +7 \\ 
Ullman Jeffrey	& 11267 & 2  & 86 & 2  & 460 & 43431 & 94.42 & 0 \\ 
Motwani Rajeev	& 9349  & 3  & 69 & 6  & 271 & 23287 & 85.93 & +3 \\ 
Fagin Ronald	& 4400  & 4  & 59 & 16 & 215 & 13604 & 63.27 & +12 \\ 
Widom Jennifer	& 4031  & 5  & 71 & 4  & 280 & 18870 & 67.39 & -1 \\ 
Florescu Daniela& 3058  & 6  & 40 & 43 & 132 & 6738  & 51.05 & +37 \\ 
Bernstein Philip& 2917  & 7  & 52 & 22 & 279 & 14721 & 52.76 & +15 \\ 
Buneman Peter	& 2001  & 8  & 43 & 39 & 158 & 6946  & 43.96 & +31 \\ 
Hellerstein Joseph& 1941  & 9  & 51 & 25 & 272 & 13212 & 48.57 & +16 \\ 
Naughton J.	& 640   & 10 & 48 & 29 & 221 & 8944  & 40.47 & +19 \\ 
Dewitt David	& 308   & 11 & 63 & 10 & 308 & 15743 & 51.11 & -1 \\ 
Koudas Nick	& 58    & 12 & 35 & 50 & 168 & 4713  & 28.05 & +38 \\ 
Sagiv Yehoshua	& -196  & 13 & 42 & 40 & 209 & 6818  & 32.62 & +27 \\ 
Chaudhuri Surajit& -278  & 14 & 41 & 41 & 239 & 7840  & 32.80 & +27 \\ 
Egenhofer Max	& -314  & 15 & 47 & 30 & 223 & 7958  & 35.69 & +15 \\ 
Livny Miron	& -597  & 16 & 61 & 12 & 310 & 14592 & 47.07 & -4 \\ 
Suciu Dan	& -659  & 17 & 54 & 19 & 285 & 11815 & 41.46 & +2 \\ 
Papadias Dimitris& -809  & 18 & 38 & 47 & 200 & 5347  & 26.73 & +29 \\ 
Lakshmanan Laks& -914  & 19 & 37 & 48 & 196 & 4969  & 25.35 & +29 \\ 
Lenzerini M.	& -1074 & 20 & 50 & 26 & 269 & 9876  & 36.71 & +6 \\ 
Abiteboul Serge& -1111 & 21 & 59 & 15 & 321 & 14347 & 44.69 & -6 \\ 
Ioannidis Yannis& -1647 & 22 & 39 & 46 & 209 & 4983  & 23.84 & +24 \\ 
Sellis Timos	& -2747 & 23 & 36 & 49 & 264 & 5461  & 20.69 & +26 \\ 
Jagadish H.	& -2924 & 24 & 52 & 24 & 303 & 10128 & 33.43 & 0 \\ 
Dayal Umeshwar	& -2975 & 25 & 44 & 34 & 306 & 8553  & 27.95 & +9 \\ 
Maier David	& -3096 & 26 & 45 & 32 & 331 & 9774  & 29.53 & +6 \\ 
Wiederhold Gio	& -3320 & 27 & 43 & 38 & 315 & 8376  & 26.59 & +11 \\ 
Ramakrishnan Raghu& -4249 & 28 & 52 & 23 & 348 & 11143 & 32.02 & -5 \\ 
Snodgrass Rick	& -4293 & 29 & 41 & 42 & 297 & 6203  & 20.89 & +13 \\ 
Srivastava Divesh& -4333 & 30 & 44 & 35 & 317 & 7679  & 24.22 & +5 \\ 
Ceri Stefano	& -4355 & 31 & 45 & 33 & 345 & 9145  & 26.51 & +2 \\ 
Kriegel Hans-Peter& -5034 & 32 & 46 & 31 & 451 & 13596 & 30.15 & -1 \\ 
Stonebraker M.	& -5643 & 33 & 62 & 11 & 380 & 14073 & 37.03 & -22 \\ 
Halevy Alon	& -5858 & 34 & 71 & 5  & 392 & 16933 & 43.20 & -29 \\ 
Abbadi Amr	& -6906 & 35 & 39 & 45 & 361 & 5652  & 15.66 & +10 \\ 
Gray Jim	& -7953 & 36 & 54 & 18 & 508 & 16563 & 32.60 & -18 \\ 
Faloutsos Christos& -8509 & 37 & 68 & 7  & 484 & 19779 & 40.87 & -30 \\ 
Jensen Christian& -8566 & 38 & 44 & 37 & 389 & 6614  & 17.00 & -1 \\ 
Agrawal Divyakant& -9199 & 39 & 40 & 44 & 433 & 6521  & 15.06 & +5 \\ 
Aalst W.	& -9811 & 40 & 48 & 28 & 468 & 10349 & 22.11 & -12 \\ 
Weikum Gerhard	& -11700& 41 & 44 & 36 & 467 & 6912  & 14.80 & -5 \\ 
Sheth Amit	& -12193& 42 & 58 & 17 & 488 & 12747 & 26.12 & -25 \\ 
Carey Michael	& -14606& 43 & 60 & 14 & 488 & 11074 & 22.69 & -29 \\ 
Franklin Michael& -14765& 44 & 60 & 13 & 559 & 15175 & 27.15 & -31 \\ 
Han Jiawei	& -15410& 45 & 77 & 3  & 653 & 28942 & 44.32 & -42 \\ 
Jajodia Sushil	& -15483& 46 & 53 & 21 & 554 & 11070 & 19.98 & -25 \\ 
Mylopoulos John& -15513& 47 & 53 & 20 & 569 & 11835 & 20.80 & -27 \\ 
Garcia-Molina Hector& -17423& 48 & 92 & 1  & 605 & 29773 & 49.21 & -47 \\ 
Bertino Elisa	& -27058& 49 & 49 & 27 & 805 & 9986  & 12.40 & -22 \\ 
Yu Philip	& -27727& 50 & 63 & 9  & 789 & 18011 & 22.83 & -41 \\ \hline
\end{tabular}
\normalsize

\end{center}
\end{table}

If we turn our attention to the bottom rows of these tables we will recognize  some 
excellent mentors, but mass producers: Elisa Bertino and Jiawei Han from databases, Georgios 
Giannakis and Jack Dongarra from the networking community, Thomas S.\ Huang, Rama Chellappa and 
Ioannis Pitas from multimedia.

\begin{table}[!t]
\caption{Rank table by $PI$ of sample ``Multimedia".}
\label{tab_rank_table_PI_s35}
\begin{center}
\footnotesize
\begin{tabular}{|l@{}||@{\hspace{.1cm}}rr@{\hspace{.1cm}}|@{\hspace{.1cm}}rr@{\hspace{.1cm}}|r|r|r|r|}\hline
\multirow{2}{*}{\bf Author } &\multicolumn{2}{@{\hspace{-.1cm}}|c@{\hspace{.1cm}}}{\bf $PI$} &\multicolumn{2}{@{\hspace{-.1cm}}|c@{\hspace{.1cm}}}{\bf $h$} & \multicolumn{1}{|c|}{\multirow{2}{*}{\bf $p$}}& \multicolumn{1}{|c|}{\multirow{2}{*}{\bf $C$}}& \multicolumn{1}{|c|}{\multirow{2}{*}{\bf $C/p$}} &\multicolumn{1}{|c@{\hspace{.1cm}}|}{change} \\
		& val   & pos& val& pos &     &      &       & $h$-$PI$\\\hline\hline
Donoho David	& 7508  & 1  & 72 & 2   & 350 & 27524& 78.64 & +1 \\ 
Cox Ingemar	& 3464  & 2  & 41 & 15  & 210 & 10393& 49.49 & +13 \\ 
Simoncelli Eero& 2619  & 3  & 47 & 12  & 227 & 11079& 48.81 & +9 \\ 
Yeo Boon-lock	& 2131  & 4  & 27 & 44  & 77  & 3481 & 45.21 & +40 \\ 
Rui Yong	& 1745  & 5  & 33 & 32  & 168 & 6200 & 36.90 & +27 \\ 
Jain Ramesh	& 1637  & 6  & 36 & 25  & 243 & 9089 & 37.40 & +19 \\ 
Yeung Minerva	& 1490  & 7  & 24 & 48  & 66  & 2498 & 37.85 & +41 \\ 
Goljan Miroslav& 1401  & 8  & 28 & 41  & 64  & 2409 & 37.64 & +33 \\ 
Wiegand Thomas	& 602   & 9  & 32 & 33  & 262 & 7962 & 30.39 & +24 \\ 
Fridrich Jessica& 472   & 10 & 27 & 45  & 118 & 2929 & 24.82 & +35 \\ 
Elad Michael	& 156   & 11 & 36 & 26  & 216 & 6636 & 30.72 & +15 \\ 
Naphade Milind	& 136   & 12 & 24 & 49  & 106 & 2104 & 19.85 & +37 \\ 
Manjunath B.	& -46   & 13 & 39 & 20  & 279 & 9314 & 33.38 & +7 \\ 
Orchard M.	& -784  & 14 & 34 & 30  & 187 & 4418 & 23.63 & +16 \\ 
Wu Min		& -1079 & 15 & 27 & 46  & 169 & 2755 & 16.30 & +31 \\ 
Li Mingjing	& -1180 & 16 & 28 & 42  & 150 & 2236 & 14.91 & +26 \\ 
Zhang Ya-Qin	& -2205 & 17 & 36 & 28  & 236 & 4995 & 21.17 & +11 \\ 
Hauptmann Alexander	& -3183 & 18 & 34 & 31  & 243 & 3923 & 16.14 & +13 \\ 
Smith John	& -3277 & 19 & 40 & 17  & 282 & 6403 & 22.71 & -2 \\ 
Zakhor Avideh	& -3468 & 20 & 38 & 23  & 268 & 5272 & 19.67 & +3 \\ 
Ebrahimi Touradj& -3520 & 21 & 31 & 37  & 272 & 3951 & 14.53 & +16 \\ 
Memon Nasir	& -3736 & 22 & 32 & 34  & 286 & 4392 & 15.36 & +12 \\ 
Li Shipeng	& -3925 & 23 & 26 & 47  & 271 & 2445 & 9.02  & +24 \\ 
Hua Xian-sheng	& -4252 & 24 & 24 & 50  & 285 & 2012 & 7.06  & +26 \\ 
Ma Wei-ying	& -4292 & 25 & 46 & 13  & 335 & 9002 & 26.87 & -12 \\ 
Ortega Antonio	& -4894 & 26 & 31 & 36  & 330 & 4375 & 13.26 & +10 \\ 
Xiong Zixiang	& -4950 & 27 & 35 & 29  & 308 & 4605 & 14.95 & +2 \\ 
Bouman C	& -5592 & 28 & 27 & 43  & 380 & 3939 & 10.37 & +15 \\ 
Wu Xiaolin	& -6332 & 29 & 31 & 39  & 337 & 3154 & 9.36  & +10 \\ 
Bovik Alan	& -8008 & 30 & 39 & 19  & 507 & 10244& 20.21 & -11 \\ 
Ramchandran Kannan& -8111 & 31 & 49 & 11  & 421 & 10117& 24.03 & -20 \\ 
Liu Bede	& -8784 & 32 & 38 & 22  & 436 & 6340 & 14.54 & -10 \\ 
Strintzis M.	& -8871 & 33 & 29 & 40  & 454 & 3454 & 7.61  & +7 \\ 
Chang Edward	& -9099 & 34 & 31 & 38  & 448 & 3828 & 8.54  & +4 \\ 
Delp Edward	& -10001& 35 & 37 & 24  & 438 & 4836 & 11.04 & -11 \\ 
Chen Liang-Gee	& -10311& 36 & 32 & 35  & 478 & 3961 & 8.29  & -1 \\ 
Tekalp A.	& -10552& 37 & 40 & 18  & 448 & 5768 & 12.88 & -19 \\ 
Unser Michael	& -10801& 38 & 54 & 6   & 465 & 11393& 24.50 & -32 \\ 
Vetterli M.	& -11139& 39 & 63 & 4   & 547 & 19353& 35.38 & -35 \\ 
Jain Anil	& -11474& 40 & 81 & 1   & 590 & 29755& 50.43 & -39 \\ 
Katsaggelos Aggelos& -11662& 41 & 36 & 27  & 504 & 5186 & 10.29 & -14 \\ 
Wang Yao	& -11705& 42 & 39 & 21  & 484 & 5650 & 11.67 & -21 \\ 
Chang Shih-Fu	& -11941& 43 & 52 & 7   & 507 & 11719& 23.11 & -36 \\ 
Nahrstedt Klara& -12286& 44 & 52 & 9   & 492 & 10594& 21.53 & -35 \\ 
Girod Bernd	& -13613& 45 & 52 & 8   & 529 & 11191& 21.16 & -37 \\ 
Pitas Ioannis	& -13849& 46 & 44 & 14  & 515 & 6875 & 13.35 & -32 \\ 
Chellappa Rama	& -14092& 47 & 50 & 10  & 604 & 13608& 22.53 & -37 \\ 
Zhang Hongjiang& -15112& 48 & 63 & 5   & 556 & 15947& 28.68 & -43 \\ 
Kuo C.		& -36848& 49 & 40 & 16  & 1148& 7472 & 6.51  & -33 \\ 
Huang Thomas	& -54047& 50 & 67 & 3   & 1172& 19988& 17.05 & -47 \\ \hline
\end{tabular}
\normalsize

\end{center}
\end{table}

But, is it really the case that only inventors and industry persons are influentials, whereas 
academia persons are mass producers? In that case, the PI index would be of little usefulness 
since the separation of influentials and mass producers would be quite straightforward. The 
answer to this question is definitely negative. From the beginning of our article we emphasized 
that this is a generic attitude of the scientists towards publishing, rather than an outcome of 
their type of careers. Thus, we can see in the ``Networks" field some academia persons such as 
Hari Balakrishnan, David Johnsonand Ion Stoica, or Peter Buneman from ``Databases" who are quite 
high in the PI ranking, even though they did not develop their careers in companies working 
with highly trained colleagues. On the other hand, P.~S.~Yu (``Databases") who spend many years 
in IBM is found at the end of Table~\ref{tab_rank_table_PI_s25}, remaining in the top-50 of
``Databases".

\section{Conclusions}
\label{sec:conclusion}

The development of indices to characterize the output of a scientist is a significant task not 
only for funding and promotion purposes, but also for discovering the scientist's ``publishing habits". 
Motivated by the question of discovering the steadily influential scientists as opposed to 
mass producers, we have defined two new areas on an scientist's citation curve:
\begin{itemize}
\item 
The {\it tail complement penalty area} (TC-area), i.e., the complement of the tail with respect 
to the line $y=h$.
\item 
the {\it ideal complement penalty area} (IC-area), i.e., the complement with respect to the 
square $p \times p$.
\end{itemize}

Using the aforementioned areas we defined two new metrics:
\begin{itemize}
\item The {\it perfectionism index based on the TC-area}, called the $PI$ index.
\item The {\it extreme perfectionism index based on the IC-area}, called the $XPI$ index.
\end{itemize}

We have performed an experimental evaluation of the behavior of the $PI$ and $XPI$ indices. 
For this purpose, we have generated three datasets (with random authors, prolific authors 
and authors with high \hi) by extracting data from the Microsoft Academic Search database. 
Our contribution is threefold:
\begin{itemize}
\item We have shown that the proposed indices are uncorrelated to previous ones, such as the \hi. 
\item We have used these new indices, in particular $PI$, to rank authors in general and, in 
      particular, to split the population of authors into two distinct groups: the ``influential" 
      ones with $PI>0$ vs. the ``mass producers" with $PI<0$.
\item Also, we have shown that ranking authors with the $PI$ index is more robust than \hi with 
      respect to self-citations, and we applied it to rank individual scientists offering some 
      explanations for the reasons behind their publishing habits.
\end{itemize}

We are already involved in the consideration of temporal issues into $PI$ by integrating 
the concepts of \chi~\cite{Sidiropoulos-Scientometrics07} into the~$PI$ index. 

\section*{Acknowledgements}
The authors wish to thank Professor Sofia Kouidou, Vice-rector of the Aristotle University of Thessaloniki, 
for stating the basic question that led to the present research.\\
The authors would also wish to thank Professor Vana Doufexi for reviewing and editing the final release of this article.\\
The offer of Microsoft to provide gratis their database API is appreciated.\\
Finally, D.Katsaros acknowledges the support of the Research Committee of the University of Thessaly through the project ``Web observatory for 
research activities in the University of Thessaly".

\bibliographystyle{plain}
\bibliography{pena}
\end{document}